\documentclass[preprint]{revtex4}
\usepackage{amsmath,amssymb}
\usepackage{graphicx}
\renewcommand{\vec}[1]{\mbox{\boldmath $#1$}}
\begin{document}
\title{Breathing oscillations and quasi-low-dimensional structures of weakly-interacting degenerate Fermi gases in highly-anisotropic traps}
\author{Takushi Nishimura}
\email[]{nishimura.takushi@ocha.ac.jp}
\affiliation{Division of Advanced Sciences, Ochadai Academic Production, Ochanomizu University, Otsuka, Bunkyo, Tokyo 112-8610, Japan}
\author{Tomoyuki Maruyama}
\email[]{maruyama.tomoyuki@nihon-u.ac.jp}
\affiliation{College of Bioresource Sciences, Nihon University, Fujisawa 252-8510, Japan}
\affiliation{Advanced Science Research Center, Japan Atomic Energy Agency, Tokai 319-1195, Japan}
\begin{abstract}
We theoretically investigate breathing oscillations of weakly-interacting degenerate Fermi gases in highly-anisotropic harmonic oscillator traps. If the traps are not highly anisotropic, the fermions behave as three-dimensional (3D) gases and exhibit the coupled breathing oscillations as studied in a previous paper~\cite{2007-01-TMTN}; Otherwise the fermions exhibit quasi-low-dimensional (QLD) properties derived from specific structures in their single-particle spectrum, called {\it QLD structures}. In the present paper, we focus on effects of the QLD structures on the breathing oscillations of the two-component fermions with symmetric population densities. Here we develop the semi-classical Thomas-Fermi approximation extended to the highly-anisotropic systems and obtain the collective frequencies in the sum-rule-scaling method and perturbation theory. As a result, we reveal that the effects of the QLD structures can not be seen in the transverse modes in the first-order perturbation and appear only in the longitudinal modes with hierarchies reflecting the QLD structures. We also demonstrate time-evolution of the oscillations in the present framework. 
\end{abstract} 
%
%
%

\maketitle
\section{Introduction \label{Sec-I}}
Development of trapping and cooling techniques of atoms yields many and various studies on trapped cold atoms, e.g., Bose-Einstein condensates (BECs)~\cite{BECrv, nobel, bec}, degenerate Fermi gases~\cite{ferG-1, ferG-2}, and mixtures of them~\cite{BferM-1, BferM-2, BferM-3, BferM-4}. In particular, the two-component Fermi gases are recently studied well in various interaction regimes. In the attraction regime, the Feshbach resonance method realizes the atomic pair condensates and BEC-BCS crossover~\cite{BCSBECe, BCSBECt}. In the strong repulsion regime, the phase separation appears~\cite{SY}. In addition, the quasi-low-dimensional (QLD) atomic gases are also realized in highly-anisotropic traps~\cite{lowdime-1, lowdime-2, lowdime-3, lowdime-4, lowdime-a1, lowdime-a2, lowdime-a3, lowdime-a4, lowdimt}. The cold atoms offer a great infrastructure for fundamental study on quantum many-particle systems. 

One of important features of the quantum many-particle systems must be collective excitations, which are often sensitive to interactions and details of quantum states. In particular, collective breathing oscillations give important diagnostic signals for properties of the trapped quantum gases in actual experiments. 

In theory, those collective excitations can be treated in the time-dependent mean-field theory for weakly-correlated systems and random-phase approximation (RPA) for small amplitude excitations~\cite{ClFer,Kry,ClSp}. Especially for pure collective excitations, simpler methods can be adopted to calculate collective frequencies of the minimal oscillations, e.g., the sum-rule method~\cite{sumF} and scaling method~\cite{scal,bohi,ToBe,2007-01-TMTN}. Note that the sum-rule and scaling methods are related to each other as explained in appendix~\ref{App-A}. Thus we call them together with a single name, {\it sum-rule-scaling method}, in the present paper. 

In a previous paper~\cite{ToBe}, one of the authors and his collaborator study spin excitations on dipole and monopole oscillations in the three-dimensional (3D) two-component Fermi gases in spherical traps using the scaling method, where the in-phase and out-of-phase oscillations are mixed, and the coupled collective frequencies reflect the phase structure of the ground state. 

In another previous paper~\cite{2007-01-TMTN}, furthermore, we study the breathing oscillations of the two-component degenerate Fermi gases in anisotropic harmonic oscillator traps with various trap frequencies and obtain the coupled collective frequencies of the longitudinal and transverse oscillations. At that time, we assume the usual trap anisotropy and apply the conventional Thomas-Fermi approximation (TFA) to the 3D gases in order to predict the collective frequencies in the sum-rule-scaling method. 

In the present paper, we extend those studies on the 3D gases to the highly-anisotropic deformed gases with the QLD properties, which gradually appear as the trap anisotropy increases. In the highly-anisotropic systems, the gases thus exhibit crossover behaviors between the 3D and QLD gases. Note that, in principle, the QLD properties should originally be given in the microscopic approach, where the many-body properties can be described in the single-particle picture with the shell-structures of the single-particle spectrum. In the present work, we thus deal with the QLD properties as appearance of the specific single-particle structures, called {\it QLD structures}, in the highly-anisotropic systems. 

The aim of the present paper is to reveal effects of the QLD properties on the breathing oscillations of the fermions in the crossover range between the 3D and QLD gases. In order to simplify the subject, we focus on the minimal oscillations of the weakly-interacting fermions with symmetric population densities, i.e., the in-phase oscillations, and calculate the collective frequencies in the sum-rule-scaling method and perturbation theory for the inter-particle interaction. In addtion, we here develop TFA to the highly-anisotropic systems in order to obtain the information of the groud state needed in the sum-rule-scaling method and to describe the QLD structures clearly. 

The contents of the present paper are as follows. In section~\ref{Sec-II}, we give theoretical explanations of the Fermi gases and breathing oscillations and aldo formulate the sum-rule-scaling method. In section~\ref{Sec-III}, we consider the ground-state properties reflecting the QLD structures by developing TFA to the highly-anisotropic systems. In section~\ref{Sec-IV}, we show calculational results of the collective frequencies and also demonstrate time-evolution of the oscillations in the present framework. In section~\ref{Sec-V}, we give a summary and outlook. 
\section{Theoretical framework \label{Sec-II}}
In this section, we give a theoretical framework for the present work. First we introduce a system of the trapped Fermi gases in subsection~\ref{Sec-II-A}. Second we explain the breathing oscillations of the gases in subsection~\ref{Sec-II-B}. Third we formulate the sum-rule-scaling method to describe the oscillations in subsection~\ref{Sec-II-C}. 
\subsection{Trapped Fermi gases \label{Sec-II-A}}
Let us consider the Fermi gases of two-component atoms with the same atomic masses, $m_{1} = m_{2} \equiv m$, where the subscripts $1$ and $2$ indicate the components of the atoms. In order to simplify the following descriptions, we take a system of unit selected as $\hbar = 1$ for the reduced Planck constant and $m = 1$ for the atomic mass. 

Assume that the atoms are simultaneously trapped in a cylindrical harmonic oscillator potential $V(\vec{r})$ denoted by 
\begin{equation}
V(\vec{r}) 
= \frac{1}{2} \left( \omega_{c}^{2} r_{c}^{2} + \omega_{z}^{2} r_{z}^{2} \right), 
\label{Eq-3}
\end{equation}
where we introduce the Cartesian coordinates, $\vec{r} = ( r_{x}, r_{y}, r_{z} )$, and $r_{c} \equiv \sqrt{r_{x}^{2} + r_{y}^{2}}$. Then the ratio of the trap frequencies, $\omega_{c} / \omega_{z}$, decides the anisotropy of this system. In the present paper, the parallel direction to the $z$-axis is called {\it longitudinal} direction, and the orthogonal directions are called {\it transverse} directions. 

The Hamiltonian $H^{(\text{tc})}$ of the two-component atoms is denoted by 
\begin{equation}
H^{(\text{tc})} 
= H_{1}^{(\text{ho})} + H_{2}^{(\text{ho})} + H_{\text{int}}^{(\text{tc})} 
\label{Eq-1}
\end{equation}
with the harmonic oscillator parts $H_{1}^{(\text{ho})}$ and $H_{2}^{(\text{ho})}$ and interaction part $H_{\text{int}}^{(\text{tc})}$ given below. Here we take the Schr\"{o}dinger representation. 

The $H_{1}^{(\text{ho})}$ and $H_{2}^{(\text{ho})}$ in eq.~(\ref{Eq-1}) are defined as 
\begin{equation}
H_{\alpha}^{(\text{ho})} 
\equiv \int d{\vec{r}}~ \Psi_{\alpha}^{\dagger}(\vec{r}) \left[ - \frac{1}{2} \nabla^{2} + V(\vec{r}) \right] \Psi_{\alpha}(\vec{r}) 
\label{Eq-1-0-1-0}
\end{equation}
for $\alpha = 1$ and $2$ with the trap potential $V(\vec{r})$ in eq.~(\ref{Eq-3}) and fermion field operators $\Psi_{\alpha}(\vec{r})$ obeying the anti-commutation relations, 
\begin{equation}
\big\{ \Psi_{\alpha}(\vec{r}), \Psi_{\beta}(\vec{s}) \big\} 
= 0 
\label{Eq-1-0-1-1}
\end{equation}
and 
\begin{equation}
\big\{ \Psi_{\alpha}(\vec{r}), \Psi_{\beta}^{\dagger}(\vec{s}) \big\} 
= \delta_{\alpha \beta} \delta(\vec{r} - \vec{s}). 
\label{Eq-1-0-1-2}
\end{equation}

The $H_{\text{int}}^{(\text{tc})}$ in eq.~(\ref{Eq-1}) for the cold atoms can be given as 
\begin{equation}
H_{\text{int}}^{(\text{tc})} 
= g \int d{\vec{r}}~ \Psi_{1}^{\dagger}(\vec{r}) \Psi_{2}^{\dagger}(\vec{r}) \Psi_{2}(\vec{r}) \Psi_{1}(\vec{r}) 
\label{Eq-2}
\end{equation}
with the contact-type pseudo-potential for the low-energy $s$-wave scattering between the $1$ and $2$ atoms, where the coupling constant $g$ is determined as $g = 4 \pi a_{1 2}$ with the $s$-wave scattering length $a_{1 2}$ in the mean-field approximation introduced below. Note that the contact-type interactions between the identical fermions must vanish because of the Pauli blocking effect. 

In order to deal with the interaction part $H_{\text{int}}^{(\text{tc})}$ in eq.~(\ref{Eq-2}), we here apply the time-dependent Hartree-Fock approximation (TDHFA) to this system~\cite{FetterWalecka}. In TDHFA, the two-body interaction in eq.~(\ref{Eq-2}) can be rewritten into a one-body interaction with the self-consistent mean-fields $\big< \Psi_{1}^{\dagger}(\vec{r}) \Psi_{1}(\vec{r}) \big>$ and $\big< \Psi_{2}^{\dagger}(\vec{r}) \Psi_{2}(\vec{r}) \big>$ including the many-body effects as mean-values. Note that the exchange term $\big< \Psi_{1}^{\dagger}(\vec{r}) \Psi_{2}(\vec{r}) \big>$ for the interaction in eq.~(\ref{Eq-2}) is neglected in TDHFA, i.e., $\big< \Psi_{1}^{\dagger}(\vec{r}) \Psi_{2}(\vec{r}) \big> = 0$, because of the assumption of the diagonal single-particle density matrix. 

In principle, TDHFA must be valid for the weakly-interacting and dilute gases, i.e., $\big| a_{1 2} \big| \ll \omega_{c}^{- 1 / 2}$ and $\big| a_{1 2} \big| \ll \omega_{z}^{- 1 / 2}$, as assumed in the present paper. If the gases are strongly correlated, e.g., in the BEC-BCS crossover~\cite{BCSBECt} or strongly-interacting QLD systems~\cite{lowdimt}, another different method may be needed to include the correlation effects beyond the present framework. 

In TDHFA, the Hamiltonian $H^{(\text{tc})}$ in eq.~(\ref{Eq-1}) for the two-component atoms can be decoupled as 
\begin{equation}
H^{(\text{tc})} 
= H_{1} + H_{2} 
\label{Eq-1-0}
\end{equation}
with the single-component Hamiltonians $H_{1}$ and $H_{2}$ denoted by 
\begin{equation}
H_{\alpha} 
= H_{\alpha}^{(\text{ho})} + H_{\alpha}^{(\text{HF})} 
\label{Eq-1-0-1}
\end{equation}
for $\alpha = 1$ and $2$, where we introduce the mean-field interaction parts, 
\begin{equation}
H_{1}^{(\text{HF})} 
\equiv g \int d{\vec{r}}~ \Psi_{1}^{\dagger}(\vec{r}) \rho_{2}^{(t)}(\vec{r}, t) \Psi_{1}(\vec{r}) - E^{(\text{HF})} 
\label{Eq-1-0-2}
\end{equation}
and 
\begin{equation}
H_{2}^{(\text{HF})} 
\equiv g \int d{\vec{r}}~ \Psi_{2}^{\dagger}(\vec{r}) \rho_{1}^{(t)}(\vec{r}, t) \Psi_{2}(\vec{r}) - E^{(\text{HF})}, 
\label{Eq-1-0-3}
\end{equation}
with half of the total interaction energy, 
\begin{equation}
E^{(\text{HF})} 
\equiv \frac{g}{2} \int d{\vec{r}}~ \rho_{1}^{(t)}(\vec{r}, t) \rho_{2}^{(t)}(\vec{r}, t), 
\label{Eq-1-0-4}
\end{equation}
for the each component of the atoms. Note that the mean-fields (or number densities), 
\begin{equation}
\rho_{\alpha}^{(t)}(\vec{r}, t) 
\equiv \big< \Psi_{\alpha}^{\dagger}(\vec{r}) \Psi_{\alpha}(\vec{r}) \big> 
\label{Eq-1-0-5}
\end{equation}
for $\alpha = 1$ and $2$, must be determined self-consistently in TDHFA. 

In the present paper, we consider the symmetric gases with the same number densities, 
\begin{equation}
\rho_{1}^{(t)}(\vec{r}, t) 
= \rho_{2}^{(t)}(\vec{r}, t) 
\equiv \rho^{(t)}(\vec{r}, t), 
\label{Eq-1-0-6}
\end{equation}
and focus on one of the atomic components, e.g., component $1$. Then we can omit the subscripts $1$ and $2$ for the atomic components according to eqs.~(\ref{Eq-1-0-1})-(\ref{Eq-1-0-6}). 

As a result, instead of $H^{(\text{tc})}$ in eq.~(\ref{Eq-1}) for the two-component atoms, we consider the Hamiltonian $H$ for the each component of the atoms denoted by 
\begin{equation}
H 
= H_{\text{ho}} + H_{\text{int}} 
\label{Eq-1-2}
\end{equation}
with the harmonic oscillator part 
\begin{equation}
H_{\text{ho}} 
\equiv \int d{\vec{r}}~ \Psi^{\dagger}(\vec{r}) \left[ - \frac{1}{2} \nabla^{2} + V(\vec{r}) \right] \Psi(\vec{r}) 
\label{Eq-1-2-1}
\end{equation}
and mean-field interaction part 
\begin{equation}
H_{\text{int}} 
\equiv g \int d{\vec{r}}~ \Psi^{\dagger}(\vec{r}) \rho^{(t)}(\vec{r}, t) \Psi(\vec{r}) - E_{\text{int}}^{(t)}, 
\label{Eq-2-2}
\end{equation}
corresponding to eq.~(\ref{Eq-1-0-1}), where the interaction energy $E_{\text{int}}^{(t)}$ corresponds to that in eq.~(\ref{Eq-1-0-4}), 
\begin{equation}
E_{\text{int}}^{(t)} 
\equiv \frac{g}{2} \int d{\vec{r}} \left[ \rho^{(t)}(\vec{r}, t) \right]^{2}. 
\label{Eq-2-2-1}
\end{equation}

Note that the results in the present paper can directly apply to the corresponding systems of the symmetric multi-component fermions beyond the two-component case because the $l$-component gases take the same formulation in eqs.~(\ref{Eq-1-2}) and (\ref{Eq-2-2}) by renormalizing the coupling constant, $g \to (l - 1) g$, where $l$ is a positive integer. 

In TDHFA, the ground state $\big| \Phi_0 \big>$ is given by the Slater determinant 
\begin{equation}
\big| \Phi_0 \big> 
= \prod_{n = 1}^{N} c_{n}^{\dagger} \big| \text{vac} \big> 
\label{Eq-4}
\end{equation}
with the fermion creation operator 
\begin{equation}
c_{n}^{\dagger} 
\equiv \int d\vec{r}~ \Psi^{\dagger}(\vec{r}) \phi_{n}(\vec{r}), 
\label{Eq-5}
\end{equation}
where the normalized single-particle wave functions $\phi_{n}(\vec{r})$ are determined by the HF equations, 
\begin{equation}
\left[ - \frac{1}{2} \nabla^{2} + V(\vec{r}) + g \rho(\vec{r}) \right] \phi_{n}(\vec{r}) 
= \varepsilon_{n} \phi_{n}(\vec{r}), 
\label{Eq-6}
\end{equation}
with the ground-state density 
\begin{equation}
\rho(\vec{r}) 
\equiv \sum_{n = 1}^{N} \big| \phi_{n}(\vec{r}) \big|^{2}. 
\label{Eq-7-0}
\end{equation}
Here the quantum number $n$ ($\ge 1$) is ordered by the single-particle energy $\varepsilon_{n}$, i.e., $\varepsilon_{m} \ge \varepsilon_{n}$ for $m > n$, and the Fermi level is determined by the particle number 
\begin{equation}
N 
= \int d{\vec{r}}~ \rho(\vec{r}) 
\label{Eq-7}
\end{equation}
owing to the normalization condition 
\begin{equation}
\int d{\vec{r}} \big| \phi_{n}(\vec{r}) \big|^{2} 
= 1. 
\label{Eq-7-1}
\end{equation}
According to eq.~(\ref{Eq-4}), the ground-state energy $E_{\text{g}} \equiv \big< \Phi_0 \big| H \big| \Phi_0 \big>$ is calculated as 
\begin{equation}
E_{\text{g}} 
= \left( \sum_{n = 1}^{N} \varepsilon_{n} \right) - E_{\text{int}} 
\label{Eq-8}
\end{equation}
with the interaction energy 
\begin{equation}
E_{\text{int}} 
\equiv \big< \Phi_0 \big| H_{\text{int}} \big| \Phi_0 \big>
= \frac{g}{2} \int d{\vec{r}} \left[ \rho(\vec{r}) \right]^{2}. 
\label{Eq-9}
\end{equation}

Lastly we comment on the validity limit of the present model. In principle, it must be valid for the weakly-interacting and dilute gases. However, if the gases are completely deformed in ultimately-anisotropic traps, the pseudo-potential in eq.~(\ref{Eq-2}) for the 3D $s$-wave scattering may not be available. At that time, we should consider more realistic inter-atomic interactions instead of the pseudo-potential. The detailed discussion is beyond the purpose of the present paper. 
\subsection{Breathing oscillations \label{Sec-II-B}}
The breathing oscillations are originally defined as compressive oscillations of fluids. Especially in spherical systems, they indicate the monopole oscillations in terms of the multipole expansion. However, in anisotropic systems, the monopole oscillations are inseparably mixed with incompressive oscillations, e.g., the quadrupole oscillations~\cite{2007-01-TMTN} and then it is necessary to redefine the breathing oscillations as mixtures of those oscillations. In the present paper, we adopt the later definition for the anisotropic systems. 

In the present work, we take the cylindrical trap in eq.~(\ref{Eq-3}) and consider cylindrical deformation in the ground and excited states. Then the breathing oscillations appear in expectation values of projected mean square radius operators 
\begin{equation}
R_{j} 
\equiv \int d{\vec{r}}~ \Psi^{\dagger}(\vec{r}) r_{j}^{2} \Psi(\vec{r}) 
\label{Eq-10}
\end{equation}
for $j = c$ and $z$. In actual experiments, those expectation values are directly observed with the absorption imaging method, and the excited gases are artificially generated with sudden change of the trap potentials. 

The minimal breathing oscillations are defined as small amplitude oscillations of variations of the expectation values of $R_{j}$ in eq.~(\ref{Eq-10}), 
\begin{equation}
s_{j} 
\equiv \big< \Phi(t) \big| R_{j} \big| \Phi(t) \big> - \big< \Phi_{0} \big| R_{j} \big| \Phi_{0} \big>, 
\label{Eq-11-0}
\end{equation}
which can be described as 
\begin{equation}
s_{j} 
=  \sum_{\nu > 0} C_{\nu} \big< \Phi_{0} \big| R_{j} \big| \Phi_{\nu} \big> e^{- i E_{\nu} t} + \text{c.c.} 
\label{Eq-11}
\end{equation}
with the time-dependent excited state 
\begin{equation}
\left| \Phi(t) \right> 
= \left| \Phi_{0} \right> + \sum_{\nu > 0} C_{\nu} \left| \Phi_{\nu} \right> e^{- i E_{\nu} t}, 
\label{Eq-12}
\end{equation}
where we introduce the excited states $\left| \Phi_{\nu} \right>$ with the excitation energy $E_{\nu}$ and minimal amplitudes $C_{\nu}$ determined by the initial condition. As shown in eq.~(\ref{Eq-11}), those oscillations directly reflect the information of the ground and excited states. 

The oscillations of $s_c(t)$ and $s_z(t)$ in eq.~(\ref{Eq-11}) are called transverse and longitudinal oscillations, respectively. In general, those oscillations are coupled and exhibit the normal mode oscillations denoted by the normal mode operators $\tilde{R}_{c}$ and $\tilde{R}_{z}$, which are given by the linear combinations of $R_{c}$ and $R_{z}$ in principle, and $\tilde{R}_{j} \approx R_{j}$ in the decoupled limit. 

Note that the monopole and quadrupole oscillations are originally defined in spherical systems, where the normal mode oscillations directly correspond to them. In general, the monopole and quadrupole oscillations indicate oscillations of $s_{m}(t) \equiv s_{c}(t) + s_{z}(t)$ and $s_{q}(t) \equiv s_{c}(t) - s_{z}(t) / 2$, respectively, which must be decoupled in spherical systems and coupled in anisotropic systems. Thus the normal mode oscillations in anisotropic systems can also be described as mixtures of the monopole and quadrupole oscillations in the similar way for the transverse and longitudinal oscillations. 

In the pure collective oscillations, the transition strength functions $\big| \big< \Phi_{0} \big| \tilde{R}_{j} \big| \Phi_{\nu} \big> \big|$ are localized in a small range of the excitation energy near the collective frequencies. Then the normal mode oscillations behave as the harmonic oscillations according to eq.~(\ref{Eq-11}), and the collective frequencies can be estimated in the sum-rule-scaling method. 
\subsection{Sum-rule-scaling method \label{Sec-II-C}}
Here we formulate the sum-rule-scaling method for the breathing oscillations, which are defined as the mixed oscillations of $s_{c}(t)$ and $s_{z}(t)$ in eq.~(\ref{Eq-11-0}) and determined from the time-dependent excited state $\big| \Phi(t) \big>$ as explained in the previous subsection. 

In the sum-rule-scaling method, the excited state $\big| \Phi(t) \big>$ is given by the scale transformation of the ground state $\big| \Phi_{0} \big>$ as 
\begin{equation}
\big| \Phi(t) \big> 
= e^{- i \xi_{\text{G}}} e^{- ( \lambda_{c} \left[ H, R_{c} \right] + \lambda_{z} \left[ H, R_{z} \right] ) / 2} \big| \Phi_{0} \big> 
\label{Eq-13}
\end{equation}
with the projected mean square radius operators $R_{c}$ and $R_{z}$ defined in eq.~(\ref{Eq-10}) and Galilei transformation factor 
\begin{equation}
e^{- i \xi_{\text{G}}} 
\equiv e^{- i ( \dot{\lambda}_{c} R_{c} + \dot{\lambda}_{z} R_{z} ) /2}, 
\label{Eq-14}
\end{equation}
where $\dot{\lambda}_{c}(t)$ and $\dot{\lambda}_{z}(t)$ represent time derivatives of $\lambda_{c}(t)$ and $\lambda_{z}(t)$, respectively. The scale parameters $\lambda_{c}(t)$ and $\lambda_{z}(t)$ in eq.~(\ref{Eq-13}) indicate the collective coordinates of the transverse and longitudinal oscillations, respectively, and are proportional to $s_{c}(t)$ and $s_{z}(t)$ in eq.~(\ref{Eq-11-0}) after all. 

In TDHFA, the ground state $\big| \Phi_{0} \big>$ is denoted by the Slater determinant in eq.~(\ref{Eq-4}), and then the excited state $\big| \Phi(t) \big>$ in eq.~(\ref{Eq-13}) can be written as 
\begin{equation}
\big| \Phi(t) \big> 
= \prod_{n = 0}^{N} d_{n}^{\dagger}(t) \big| \text{vac} \big> 
\label{Eq-15}
\end{equation}
with 
\begin{equation}
d_{n}^{\dagger}(t) 
\equiv \int d{\vec{r}}~ \Psi^{\dagger}(\vec{r}) \tilde{\phi}_{n}(\vec{r}, t), 
\label{Eq-16}
\end{equation}
where we introduce the scaled wave functions defined as 
\begin{equation}
\tilde{\phi}_{n}(\vec{r}, t) 
\equiv e^{- i ( \dot{\lambda}_{c} r_{c}^{2} + \dot{\lambda}_{z} r_{z}^{2} ) / 2} e^{\lambda_{c} + \lambda_{z} / 2} \phi_{n}\left( \vec{\tilde{r}}_{c}, \tilde{r}_{z} \right) 
\label{Eq-17}
\end{equation}
with $\vec{\tilde{r}}_{c} \equiv e^{\lambda_{c}} \vec{r}_{c} \equiv \big( e^{\lambda_{c}} r_{x}, e^{\lambda_{c}} r_{y} \big)$ and $\tilde{r}_{z} \equiv e^{\lambda_{z}} r_{z}$. The scaled wave functions $\tilde{\phi}_{n}(\vec{r}, t)$ in eq.~(\ref{Eq-17}) satisfy the normalization and continuity conditions. 

The time-dependent variational principle, 
\begin{equation}
\delta \int d{t}~ \mathcal{L}\big[ \lambda, \dot{\lambda} \big] 
= 0, 
\label{Eq-19}
\end{equation}
gives equations of motion for the scale parameters $\lambda_{c}(t)$ and $\lambda_{z}(t)$ in eq.~(\ref{Eq-13}), where we define the Lagrangian $\mathcal{L}$ as 
\begin{equation}
\mathcal{L}\big[ \lambda, \dot{\lambda} \big] 
\equiv \big< \Phi(t) \big| i \frac{d}{dt} - H \big| \Phi(t) \big>. 
\label{Eq-19-1}
\end{equation}
Note that the sum-rule-scaling method must be in TDHFA in the present framework because the variational space in eq.~(\ref{Eq-19}) is limited in the configuration space of the Slater determinants according to eq.~(\ref{Eq-15}). 

By substituting eq.~(\ref{Eq-15}) into eq.~(\ref{Eq-19-1}), $\mathcal{L}\big[ \lambda, \dot{\lambda} \big]$ in eq.~(\ref{Eq-19-1}) can be described as 
\begin{equation}
\mathcal{L}\big[ \lambda, \dot{\lambda} \big] 
= \sum_{j = c, z} \mathcal{M}_{j}^{(\lambda)} \dot{\lambda}_{j}^{2} - \mathcal{H}, 
\label{Eq-20}
\end{equation}
where we define the mass parameters 
\begin{equation}
\mathcal{M}_{j}^{(\lambda)} 
\equiv e^{- 2 \lambda_{j}} \mathcal{M}_{j} 
\label{Eq-21}
\end{equation}
with 
\begin{equation}
\mathcal{M}_{j} 
\equiv \int d{\vec{r}}~ r_{j}^{2} \rho(\vec{r}) 
\label{Eq-22}
\end{equation}
and excitation energy 
\begin{equation}
\mathcal{H}\big[ \lambda, \dot{\lambda} \big] 
\equiv \big< \Phi(t) \big| H \big| \Phi(t) \big> - E_{\text{g}} 
\label{Eq-23}
\end{equation}
with the ground energy $E_{\text{g}}$ in eq.~(\ref{Eq-8}). 

The excitation energy $\mathcal{H}\big[ \lambda, \dot{\lambda} \big]$ in eq.~(\ref{Eq-23}) is calculated as 
\begin{equation}
\mathcal{H}\big[ \lambda, \dot{\lambda} \big] 
= \frac{1}{2} \sum_{j = c, z} \mathcal{M}_{j}^{(\lambda)} \dot{\lambda}_{j}^{2} + \mathcal{V}\big[ \lambda \big] 
\label{Eq-24}
\end{equation}
with the potential parameter 
\begin{equation}
\mathcal{V}\big[ \lambda \big] 
\equiv \sum_{j = c, z} \left( e^{2 \lambda_{j}} K_{j} + e^{- 2 \lambda_{j}} U_{j} \right) + e^{2 \lambda_{c} + \lambda_{z}} E_{\text{int}}, 
\label{Eq-25}
\end{equation}
where $K_{j}$ and $U_{j}$ indicate the kinetic and trap-potential energies of the ground state in the $j$-direction motion, 
\begin{equation}
K_{j} 
\equiv \int d{\vec{r}} \big< \Phi_{0} \big| \Psi^{\dagger}(\vec{r}) \left( - \frac{1}{2} \nabla_{j}^{2} \right) \Psi(\vec{r}) \big| \Phi_{0} \big> 
\label{Eq-25-1}
\end{equation}
and 
\begin{equation}
U_{j} 
\equiv \int d{\vec{r}} \big< \Phi_{0} \big| \Psi^{\dagger}(\vec{r}) \left( \frac{1}{2} \omega_{j}^2 r_{j}^{2} \right) \Psi(\vec{r}) \big| \Phi_{0} \big>. 
\label{Eq-25-2}
\end{equation}

In order to deal with the minimal oscillations, $\big| \lambda_{c} \big| \ll 1$ and $\big| \lambda_{z} \big| \ll 1$, studied in the present work, we expand the excitation energy $\mathcal{H}\big[ \lambda, \dot{\lambda} \big]$ in eq.~(\ref{Eq-23}) by $\lambda_{c}$ and $\lambda_{z}$ and take the terms up to the second order, 
\begin{equation}
\mathcal{H}\big[ \lambda, \dot{\lambda} \big] 
\simeq \frac{1}{2} \left( \dot{\vec{\lambda}}^{T} B \dot{\vec{\lambda}} + \vec{\lambda}^{T} C \vec{\lambda} \right) 
\label{Eq-26}
\end{equation}
with $\vec{\lambda} \equiv \big( \lambda_{c}, \lambda_{z} \big)^{T}$, where the superscript $T$ indicates the transposition. The mass parameter matrix $B$ in eq.~(\ref{Eq-26}) is defined as 
\begin{equation}
B 
\equiv \left[ 
\begin{array}{cc} 
\mathcal{M}_{c} & 0 
\\
0 & \mathcal{M}_{z} 
\end{array} 
\right], 
\label{Eq-27}
\end{equation}
where the diagonalization of $B$ is due to $\big[ \big[ H, R_{c} \big], R_{z} \big] = 0$ and $\big[ \big[ H, R_{z} \big], R_{c} \big] = 0$. The restoring force matrix $C$ in eq.~(\ref{Eq-26}) is defined as 
\begin{equation}
C 
\equiv \left[ 
\begin{array}{cc} 
4 \left( K_{c} + U_{c} \right) + 4 E_{\text{int}} 
& 
2 E_{\text{int}} 
\\
2 E_{\text{int}} 
& 
4 \left( K_{z} + U_{z} \right) + E_{\text{int}} 
\end{array} 
\right]. 
\label{Eq-28}
\end{equation}

Note that the first order term of $\vec{\lambda}$ in eq.~(\ref{Eq-26}) must vanish because of the generalized virial theorem, 
\begin{equation}
K_{c} 
= U_{c} - E_{\text{int}} 
\label{Eq-29}
\end{equation}
and
\begin{equation}
K_{z} 
= U_{z} - \frac{1}{2} E_{\text{int}}. 
\label{Eq-30}
\end{equation}

In order to diagonalize $\mathcal{H}\big[ \lambda, \dot{\lambda} \big]$ in eq.~(\ref{Eq-26}), we introduce a complete set of orthonormal eigenvectors, 
\begin{equation}
\left( B^{- 1 / 2} C B^{- 1 / 2} \right) \vec{\eta}_{j} 
= \Omega_{j}^{2} \vec{\eta}_{j} 
\label{Eq-32}
\end{equation}
for $j = c$ and $z$, where the eigenvalues $\Omega_{j}^{2}$ must be positive for the stable ground state. Then $\vec{\lambda}$ in eq.~(\ref{Eq-26}) can be described as 
\begin{equation}
\vec{\lambda} 
= B^{- 1 / 2} \sum_{j = c, z} u_{j} \vec{\eta}_{j} 
\label{Eq-33}
\end{equation}
with 
\begin{equation}
u_{j} 
= \vec{\eta}_{j}^{T} B^{1 / 2} \vec{\lambda}. 
\label{Eq-34}
\end{equation}

By substituting eq.~(\ref{Eq-33}) into eq.~(\ref{Eq-26}), $\mathcal{H}\big[ \lambda, \dot{\lambda} \big]$ in eq.~(\ref{Eq-26}) can be written as the diagonal form, 
\begin{equation}
\mathcal{H}\big[ \lambda, \dot{\lambda} \big] 
= \frac{1}{2} \left( \dot{\tilde{\vec{\lambda}}}^{T} \dot{\tilde{\vec{\lambda}}} + \tilde{\vec{\lambda}}^{T} \tilde{C} \tilde{\vec{\lambda}} \right) 
\label{Eq-35}
\end{equation}
with the normal mode vector $\tilde{\vec{\lambda}} \equiv (u_{c}, u_{z})^{T}$ and eigenvalue matrix 
\begin{equation}
\tilde{C} 
\equiv \left[ 
\begin{array}{cc} 
\Omega_{c}^{2} 
& 
0 
\\
0 
& 
\Omega_{z}^{2} 
\end{array} 
\right]. 
\label{Eq-36}
\end{equation}
According to eqs.~(\ref{Eq-35}) and (\ref{Eq-36}), $\Omega_{c}$ and $\Omega_{z}$ in eq.~(\ref{Eq-32}) correspond to the collective frequencies of the transverse and longitudinal normal modes, respectively. 

As a result, we can rewrite the coupled scale transformation in eq.~(\ref{Eq-13}) into the decoupled form, 
\begin{equation}
\big| \Phi(t) \big> 
= e^{- i \xi_{\text{G}}} e^{- ( u_{c} \left[ H, \tilde{R}_{c} \right] + u_{z} \left[ H, \tilde{R}_{z} \right] ) / 2} \big| \Phi_{0} \big> 
\label{Eq-37}
\end{equation}
with 
\begin{equation}
e^{- i \xi_{\text{G}}} 
= e^{- i ( \dot{u}_{c} \tilde{R}_{c} + \dot{u}_{z} \tilde{R}_{z} ) /2}, 
\label{Eq-37-2}
\end{equation}
where we introduce the normal mode operators 
\begin{equation}
\tilde{R}_{j} 
\equiv \vec{\eta}_{j}^{T} B^{- 1 / 2} \vec{R} 
\label{Eq-38}
\end{equation}
with $\vec{R} \equiv \big( R_{c}, R_{z} \big)^{T}$. 

Here we comment on the relationship between the sum-rule and scaling methods. In the above formulation, the both methods predict the same collective frequencies as explained in appendix~\ref{App-A}. However, in principle, they are originally different approaches playing the complementary roles in the study on the collective oscillations. In fact, the scaling method clearly gives the time-dependent excitation state in eq.~(\ref{Eq-13}) and determine the normal mode operators $\tilde{R}_{j}$ in eq.~(\ref{Eq-38}), differently from the sum-rule method; On the other hand, the sum-rule method can give more detailed information of the oscillations, e.g., collectivity of the oscillations, with additional data of the energy moments, differently from the scaling method. 

In conclusion, the breathing oscillations depend only on the parameters $\mathcal{M}_{c}$, $\mathcal{M}_{z}$, $K_{c}$, $K_{z}$, $U_{c}$, $U_{z}$, and $E_{\text{int}}$ determined from the ground state properties. Actually the collective frequencies are calculated in eq.~(\ref{Eq-32}) with these parameters, and the precise behaviors are predicted via the time-dependent excitation state in eq.~(\ref{Eq-13}) by solving the classical dynamics denoted by eqs.~(\ref{Eq-20}) and (\ref{Eq-26}). 
\section{Ground state properties \label{Sec-III}}
In this section, we calculate the parameters needed to predict the breathing oscillations in the sum-rule-scaling method. The parameters are determined only from the ground state properties as explained in the previous section. The collective frequencies must thus reflect the QLD structures through the properties. Here we consider the ground state with the QLD structures by developing TFA to the highly-anisotropic systems in subsection~\ref{Sec-III-A}, and then determine the parameters in the present approach in subsection~\ref{Sec-III-B}. 
\subsection{QLD structures and extended TFA \label{Sec-III-A}}
The QLD structures are defined as the specific single-particle structures appeared in the highly-anisotropic systems. If $\omega_{c} \ll \omega_{z}$, the system exhibits the quasi-two-dimensional (Q2D) structure related to the Q2D system; If $\omega_{c} \gg \omega_{z}$, the system exhibits the quasi-one-dimensional (Q1D) structure related to the Q1D system. In the present paper, we treat the both cases. 

In general, definition of the QLD systems can not be unique. That is because the QLD systems are artificially defined with the handed variable-separation for the anisotropic directions, although the inter-particle interactions can not permit the variable-separation. In other words, the QLD systems merely indicate the partial systems chosen by hand. In principle, such a partial system must have effective interactions owing to the interactions between the partial and residual systems because physics must not depend on the selection of the partial system and must be described in the total system. 

The QLD structures are fundamentally independent of the artificial definition of the QLD systems and, however, must be related to the QLD systems in the highly-anisotropic and weakly-interacting limits. As demonstrated below, the relationship can clearly be seen in the perturbative approach started from the non-interacting asymptotic state with the exact variable-separation. 

In the first-order perturbation theory, all physical values are evaluated from the non-perturbative state, i.e., the ground state of the non-interacting gases. In the state, the single-particle energies in eq.~(\ref{Eq-6}) are given as 
\begin{equation}
\varepsilon_{n} 
= \sum_{j = x, y, z} \omega_{j} \left( n_{j} + \frac{1}{2} \right) 
\equiv \sum_{j = x, y, z} e_{j} 
\label{Eq-42}
\end{equation}
with $n = \{ n_{x}, n_{y}, n_{z} \}$ and $\omega_{x} = \omega_{y} \equiv \omega_{c}$. The corresponding single-particle wave-functions, $\phi_{n}(\vec{r})$, are denoted by the variable separation form, 
\begin{equation}
\phi_{n}{\left( \vec{r} \right)} 
= \prod_{j = x, y, z} \varphi_{j}{\left( r_{j} \right)} 
\label{Eq-43}
\end{equation}
with 
\begin{equation}
\varphi_{j}{\left( r_{j} \right)} 
\equiv \frac{\omega_{j}^{1 / 4}}{\pi^{1 / 4}} \sqrt{\frac{e^{- \omega_{j} r_{j}^{2}}}{2^{n_{j}} n_{j}!}} H_{n_{j}}{\left( \sqrt{\omega_{j}} r_{j} \right)}. 
\label{Eq-44}
\end{equation}
Here $H_{n}(x)$ indicates the Hermite functions. 

In order to develop TFA, we here introduce the Wigner function defined as 
\begin{equation}
f(\vec{r}, \vec{p}) 
\equiv \int d{\vec{s}}~ e^{- i \vec{p} \cdot \vec{s}} \left< \Psi^{\dagger}{\left( \vec{r} - \frac{\vec{s}}{2} \right)} \Psi{\left( \vec{r} + \frac{\vec{s}}{2} \right)} \right>. 
\label{Eq-45}
\end{equation}
Then the expectation value of an arbitrary one-body operator 
\begin{equation}
A 
\equiv \int d{\vec{r}}~ \Psi^{\dagger}{\left( \vec{r} \right)} a{\left( \hat{\vec{r}}, \hat{\vec{p}} \right)} \Psi{\left( \vec{r} \right)} 
\label{Eq-46}
\end{equation}
can be written as 
\begin{equation}
\big< A \big> 
= \iint \frac{d{\vec{r}} d{\vec{p}}}{(2 \pi)^{3}}~ a^{(\text{cl})}(\vec{r}, \vec{p}) f(\vec{r}, \vec{p}) 
\label{Eq-47}
\end{equation}
with the Weyl classical function $a^{(\text{cl})}(\vec{r}, \vec{p})$ corresponding to the quantum mechanical operator $a(\hat{\vec{r}}, \hat{\vec{p}})$. 

In the non-interacting gases, the Wigner function $f(\vec{r}, \vec{p})$ in eq.~(\ref{Eq-45}) can be described as 
\begin{equation}
f(\vec{r}, \vec{p}) 
= \sum_{n} \mathcal{F}(\varepsilon_{n}) \prod_{j = x, y, z} f_{j}(r_{j}, p_{j}; e_{j}) 
\label{Eq-48}
\end{equation}
with the zero-temperature Fermi distribution function 
\begin{equation}
\mathcal{F}(\varepsilon_{n}) 
\equiv \Theta{\left( e_{\text{F}} - \varepsilon_{n} \right)} 
\label{Eq-49-2}
\end{equation}
and variable separation parts 
\begin{equation}
f_{j}(r_{j}, p_{j}; e_{j}) 
\equiv \int_{- \infty}^{\infty} d{s}~ e^{-i p_{j} s} \varphi_{j}^{*}{\left( r_{j} - \frac{s}{2} \right)} \varphi_{j}{\left( r_{j} + \frac{s}{2} \right)} 
\label{Eq-49}
\end{equation}
according to eqs.~(\ref{Eq-42}) and (\ref{Eq-43}), where we introduce the Fermi energy $e_{\text{F}}$ and Heaviside step function $\Theta{\left( x \right)}$ obeying $\Theta{\left( x > 0 \right)} = 1$ and $\Theta{\left( x < 0 \right)} = 0$. The Wigner function $f(\vec{r}, \vec{p})$ in eq.~(\ref{Eq-48}) is simplified in TFA by smoothing the energy-level density as explained below. 
\subsubsection{Conventional TFA \label{Sec-III-A-1}}
The conventional 3DTFA, explained in Appendix~\ref{App-B}, must be valid only if the traps are not highly anisotropic; Otherwise 3DTFA loses the validity because of the QLD structures. In fact, it is clearly seen in the single-particle energy-level densities~\cite{Nilsson}. Here we show them in the oblate and prolate deformed gases in order. 

First we consider the oblate deformed gases, $\omega_{z} > \omega_{c}$, and take the anisotropy $a_{z} \equiv \omega_{z} / \omega_{c}$ to be an integer in order to simplify the following descriptions, where the single-particle energies $\varepsilon_{n}$ in eq.~(\ref{Eq-42}) can be written as 
\begin{equation}
\varepsilon_{n} 
= \omega_{c} q_{c} + e_{0} 
\equiv e_{q} 
\label{Eq-50-0}
\end{equation}
with the quantum number 
\begin{equation}
q_{c} 
\equiv n_{x} + n_{y} + a_{z} n_{z} 
\label{Eq-50-0-1}
\end{equation}
for the $c$-direction motion and zero point energy 
\begin{equation}
e_{0} 
\equiv \omega_{c} + \frac{1}{2} \omega_{z}. 
\label{Eq-50-0-2}
\end{equation}

At that time, we can perform coarse graining for the energy width $\omega_{c}$ in eq.~(\ref{Eq-50-0}) and introduce the discrete energy-level density $D(e_{q})$ defined as the mean number density of the energy-levels in the energy width $\omega_{c}$ around the energy eigenvalue $e_{q}$ instead of the density of states in eq.~(\ref{Eq-B3}). 

As a result, we obtain 
\begin{equation}
D(e_{q}) 
= \frac{(q_{z} + 1) (2 q_{c} + 2 - a_{z} q_{z})}{2 \omega_{c}} 
\label{Eq-50}
\end{equation}
with the maximum quantum number 
\begin{equation}
q_{z} 
\equiv \text{int}{\left[ \frac{e_{q} - e_{0}}{\omega_{z}} \right]} 
\label{Eq-50-1}
\end{equation}
for the $z$-direction motion, where the round-down function $\text{int}{\left[ x \right]}$ returns the integer part of the real number $x$. 

If the trap anisotropy $a_{z}$ is not so large, most of the fermions occupy the energy levels in the range denoted by $e_{q} \gg \omega_{z} > \omega_{c}$ because of the Pauli exclusion principle. Then we can take the smoothing scheme for both of the quantum numbers, $q_{c} \approx ( \varepsilon - e_{0} ) / \omega_{c}$ and $q_{z} \approx ( \varepsilon - e_{0} ) / \omega_{z}$, to obtain the smoothed energy level density $D(\varepsilon)$ from the discrete energy level density $D(e_{q})$ in eq.~(\ref{Eq-50}). As a result, we obtain 
\begin{equation}
D(\varepsilon) 
\approx D^{(3)}(\varepsilon) 
= \frac{\varepsilon^{2}}{2 \omega_{c}^{2} \omega_{z}} \Theta{\left( \varepsilon \right)}, 
\label{Eq-51}
\end{equation}
where $D^{(3)}(\varepsilon)$ indicates the energy level density in 3DTFA as described in eq.~(\ref{Eq-B7}). 

In the highly-anisotropic Q2D limit, most of the fermions occupy the energy levels in the range denoted by $\omega_{z} > e_{q} \gg \omega_{c}$, i.e., $q_{z} = 0$. Then we take the smoothing scheme only for $q_{c} \approx ( \varepsilon - e_{0} ) / \omega_{c}$ and obtain 
\begin{equation}
D(\varepsilon) 
\approx D^{(2)}(\varepsilon) 
= \frac{1}{\omega_{c}^{2}} \left( \varepsilon - \frac{\omega_{z}}{2} \right) \Theta{\left( \varepsilon - \frac{\omega_{z}}{2} \right)}, 
\label{Eq-52}
\end{equation}
where $D^{(2)}(\varepsilon)$ indicates the energy level density in 2DTFA as described in eq.~(\ref{Eq-B7}). 

\begin{figure}[ht]
  \begin{center}
    \begin{tabular}{c}
       \includegraphics[scale=0.5,angle=-90]{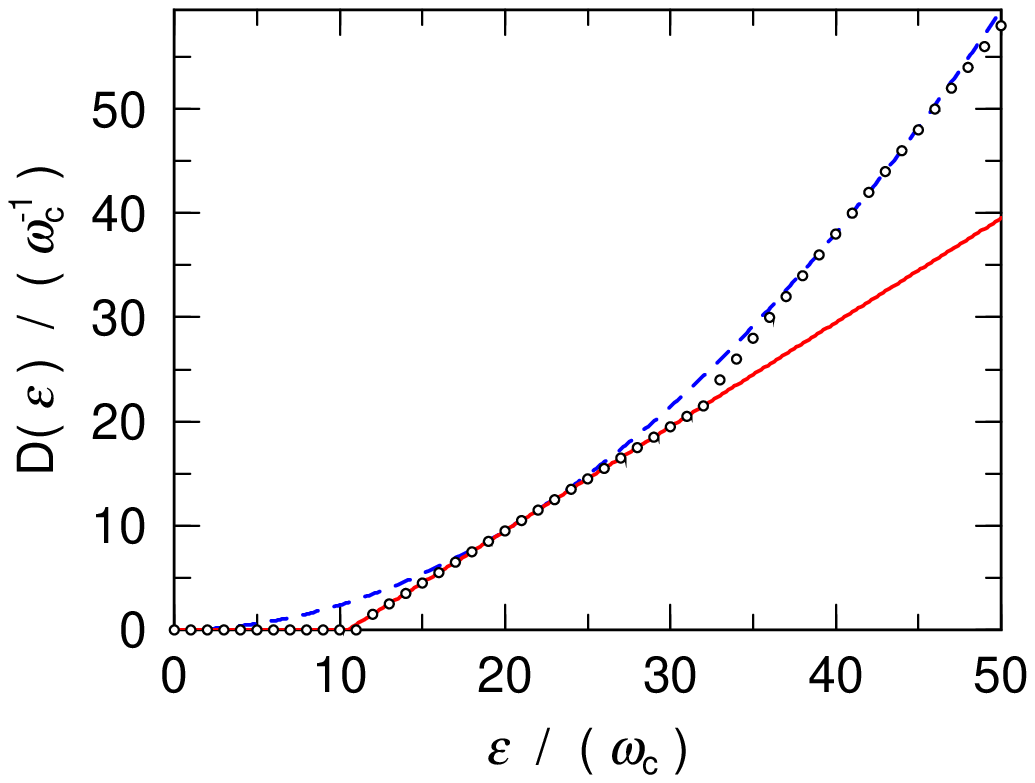}
    \end{tabular}
  \end{center}
\caption{(Color online) 
The energy level densities when $\omega_{z} / \omega_{c} = 21$. The open circles represent the values of $D(e_{q})$. The dashed and solid lines represent the results in 3DTFA and 2DTFA, respectively. 
}
\label{Fig-dos-ob}
\end{figure}

In fig.~\ref{Fig-dos-ob}, we show the energy level densities when $\omega_{z} / \omega_{c} = 21$. If $\varepsilon \gg \omega_{z}$, the values of $D(e_{q})$ (denoted by the open circles) approximately agree with those of $D^{(3)}(\varepsilon)$ (denoted by the dashed line) as described in eq.~(\ref{Eq-51}); Otherwise the both results exhibit the visible difference reflecting the QLD structures. If $\varepsilon < 31.5 \omega_{c} = 1.5 \omega_{z}$, i.e., $q_{z} = 0$, the values of $D(e_{q})$ agree with those of $D^{(2)}(\varepsilon)$ (denoted by the solid line) as described in eq.~(\ref{Eq-52}). 

Second we consider the prolate deformed gases, $\omega_{c} > \omega_{z}$, and also take the anisotropy $a_{c} \equiv \omega_{c} / \omega_{z}$ to be an integer, where we can apply the same way in the oblate deformed gases to obtain the energy-level density $D(e_{q})$ with exchange of the subscripts, $c \leftrightarrow z$, and obtain 
\begin{equation}
D(e_{q}) 
= \frac{(q_{c} + 1) (q_{c} + 2)}{2 \omega_{z}}. 
\label{Eq-53}
\end{equation}

According to the same way in the oblate deformed gases, the energy level density $D(e_{q})$ in eq.~(\ref{Eq-53}) agrees with that in 3DTFA, $D^{(3)}(\varepsilon)$, as described in eq.~(\ref{Eq-51}) if the trap anisotropy $a_{c}$ is not so large; Otherwise they disagree because of the QLD structures. Especially in the highly-anisotropic Q1D limit, $\omega_{c} > e_{q} \gg \omega_{z}$ and $q_{c} = 0$, we can take the smoothing scheme only for $q_{z} \approx ( \varepsilon - e_{0} ) / \omega_{z}$ and obtain 
\begin{equation}
D(\varepsilon) 
\approx D^{(1)}(\varepsilon) 
= \frac{1}{\omega_{z}} \Theta{\left( \varepsilon - \omega_{c} \right)}, 
\label{Eq-54}
\end{equation}
where $D^{(1)}(\varepsilon)$ indicates the energy level density in 1DTFA as described in eq.~(\ref{Eq-B7}). 

\begin{figure}[ht]
  \begin{center}
    \begin{tabular}{c}
       \includegraphics[scale=0.5,angle=-90]{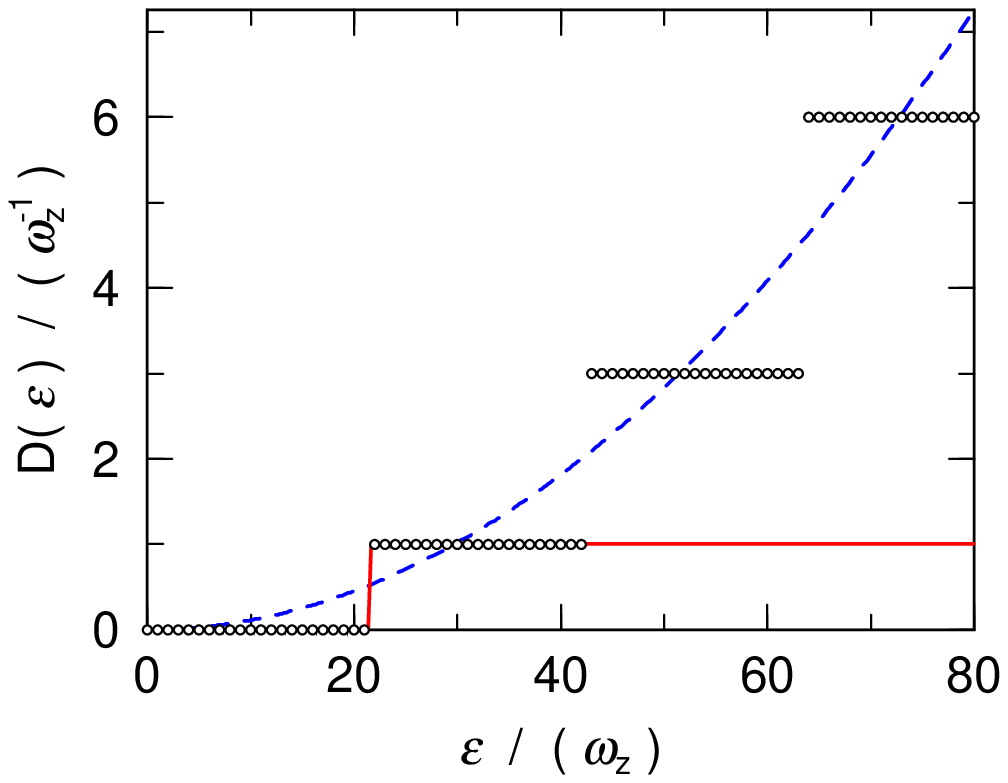}
    \end{tabular}
  \end{center}
\caption{(Color online) 
The energy level densities when $\omega_{c} / \omega_{z} = 21$. The open circles represent the values of $D(e_{q})$. The dashed and solid lines represent the results in 3DTFA and 1DTFA, respectively. 
}
\label{Fig-dos-pr}
\end{figure}

In fig.~\ref{Fig-dos-pr}, we show the energy level densities when $\omega_{c} / \omega_{z} = 21$. If $\varepsilon \gg \omega_{c}$, the values of $D(e_{q})$ (denoted by the open circles) approximately agree with those of $D^{(3)}(\varepsilon)$ (denoted by the dashed line) in analogy with the case of the oblate deformed gases. If $\varepsilon < 42 \omega_{z} = 2 \omega_{c}$, i.e.,  $q_c = 0$, the values of $D(e_{q})$ agree with those of $D^{(1)}(\varepsilon)$ (denoted by the solid line) as described in eq.~(\ref{Eq-54}). The appearance of the QLD structures in the prolate deformed gases in fig.~\ref{Fig-dos-pr} is clearer than that in the oblate deformed gases in fig.~\ref{Fig-dos-ob}. 

Note that, as shown in figs.~\ref{Fig-dos-ob} and \ref{Fig-dos-pr}, the QLD structures exhibit the hierarchic structures associated with the maximum quantum numbers for the narrow direction motions, i.e., $q_{z}$ for the oblate deformed gases and $q_{c}$ for the prolate deformed gases. The hierarchic structures must reflect the specific feature of the QLD structures and are also shown in the following results. In general, if we expand the field operator $\Psi(\vec{r})$ with the complete set of the single-particle wave-functions $\phi_{n}(\vec{r})$ in eq.~(\ref{Eq-43}) with the variable separation form, the expanded terms are denoted by two types of the quantum numbers, i.e., the fine and coarse modes, and the coarse mode produces the hierarchic structures in the highly-anisotropic systems in principle. After all, the QLD structures determine the fundamental mechanism of the appearance of the QLD gases and dimensionality. 
\subsubsection{Extended TFA \label{Sec-III-A-2}}
As shown in the energy-level densities described above, the conventional 3DTFA should not be applied to the highly-anisotropic systems because of the QLD structures. Here we introduce another approach to develop TFA to the highly-anisotropic systems. 

In order to deal with the QLD structures, we smooth the Wigner function $f(\vec{r}, \vec{p})$ in eq.~(\ref{Eq-48}) only for the $c$-direction in the oblate deformed gases and only for the $z$-direction in the prolate deformed gases. Then we obtain 
\begin{equation}
f(\vec{r}, \vec{p}) 
= \int_{0}^{\infty} d{\varepsilon}~ G^{(j)}(\varepsilon) 
\label{Eq-55}
\end{equation}
for $j = 2$ and $1$, where we introduce 
\begin{eqnarray}
&& 
G^{(2)}(\varepsilon) 
\equiv \mathcal{F}{\left[ \varepsilon + e_{\text{cl}}^{(2)}(\vec{r}_{c}, \vec{p}_{c}) \right]} 
\nonumber \\ && \times 
\sum_{n_{z}} \delta{\left( \varepsilon - e_{z} \right)} f_{z}(r_{z}, p_{z}; e_{z}) 
\label{Eq-56}
\end{eqnarray}
for the oblate deformed gases with the Q2D classical energy 
\begin{equation}
e_{\text{cl}}^{(2)}(\vec{r}_{c}, \vec{p}_{c}) 
\equiv \frac{1}{2} p_{c}^{2} + \frac{1}{2} \omega_{c}^{2} r_{c}^{2} 
\label{Eq-56-2}
\end{equation}
and 
\begin{eqnarray}
&& 
G^{(1)}(\varepsilon) 
\equiv \mathcal{F}{\left[ \varepsilon + e_{\text{cl}}^{(1)}(r_{z}, p_{z}) \right]} 
\nonumber \\ && \times 
\sum_{n_{x}, n_{y}} \delta{\left( \varepsilon - e_{x} - e_{y} \right)} \prod_{j = x, y} f_{j}(r_{j}, p_{j}; e_{j}) 
\label{Eq-57}
\end{eqnarray}
in the prolate deformed gases with the Q1D classical energy 
\begin{equation}
e_{\text{cl}}^{(1)}(r_{z}, p_{z}) 
\equiv \frac{1}{2} p_{z}^{2} + \frac{1}{2} \omega_{z}^{2} r_{z}^{2}. 
\label{Eq-57-2}
\end{equation}

Note that, if $\varepsilon \gg \omega_{c}$ and $\varepsilon \gg \omega_{z}$, the Wigner function $f(\vec{r}, \vec{p})$ in eq.~(\ref{Eq-55}) agrees with that in 3DTFA, 
\begin{equation}
f^{(3)}(\vec{r}, \vec{p}) 
\equiv \mathcal{F}{\left[ e_{\text{cl}}(\vec{r}, \vec{p}) \right]}, 
\label{Eq-58-0}
\end{equation}
according to the smoothing scheme, where we introduce the 3D classical energy 
\begin{equation}
e_{\text{cl}}(\vec{r}, \vec{p}) 
\equiv \frac{1}{2} p^{2} + \frac{1}{2} \left( \omega_{c}^{2} r_{c}^{2} + \omega_{z}^{2} r_{z}^{2} \right). 
\label{Eq-58-2}
\end{equation}

Thus we can rewrite the Wigner function $f(\vec{r}, \vec{p})$ in eq.~(\ref{Eq-55}) as 
\begin{eqnarray}
f(\vec{r}, \vec{p}) 
&=& \int_{0}^{\varepsilon_{\text{c}}} d{\varepsilon}~ G^{(j)}(\varepsilon) 
\nonumber \\ 
&+& f^{(3)}(\vec{r}, \vec{p}) \Theta{\left[ e_{\text{cl}}(\vec{r}, \vec{p}) - \varepsilon_{\text{c}} \right]} 
\label{Eq-58}
\end{eqnarray}
for the oblate ($j = 2$) and prolate ($j = 1$) deformed gases with a large cutoff energy $\varepsilon_{\text{c}}$. The rewritten formula in eq.~(\ref{Eq-58}) is called {\it extended TFA} (ETFA) in the present paper. If $\varepsilon_{\text{c}} > e_{\text{F}}$, ETFA exactly reproduces the original Wigner function $f(\vec{r}, \vec{p})$ in eq.~(\ref{Eq-55}). 

The QLD and hierarchic structures can be seen in eq.~(\ref{Eq-58}) through eqs.~(\ref{Eq-56}) and (\ref{Eq-57}), and thus ETFA reproduces the QLD structures below the cutoff energy $\varepsilon_{\text{c}}$; On the other hand, ETFA does not reproduce the QLD structures above the cutoff energy $\varepsilon_{\text{c}}$ as shown in eq.~(\ref{Eq-58-0}). In ETFA, the value of the cutoff energy $\varepsilon_{\text{c}}$ determines the contribution of the QLD structures, and we can manually select the value of $\varepsilon_{\text{c}}$ to see the detail of the contribution. 

Note that, pragmatically, it is not necessary to include the QLD structures in the whole range of the single-particle energy $\varepsilon$ because the contribution of the QLD structures is principally derived from the low-energy parts of the structures. In fact, the results in ETFA with a finite value of the cutoff energy $\varepsilon_{\text{c}}$ can be in good agreement with those in the exact calculations, i.e., in the $\varepsilon_{\text{c}} \to \infty$ limit, as demonstrated in the next section. 
\subsubsection{Fermi energy and crossover behaviors \label{Sec-III-A-3}}
In order to see the contribution of the QLD structures in eq.~(\ref{Eq-55}) in the highly-anisotropic systems, we show the Fermi energy $e_{\text{F}}$ for an example. As a typical behavior in the highly-anisotropic systems, the Fermi energy exhibits the crossover behaviors between the 3D and QLD systems with the hierarchic structures in ETFA. 

The Fermi energy $e_{\text{F}}$ is determined by the particle number condition, 
\begin{equation}
N 
= \iint \frac{d{\vec{r}} d{\vec{p}}}{(2 \pi)^{3}}~ f(\vec{r}, \vec{p}) 
\label{Eq-60}
\end{equation}
with the Wigner function $f(\vec{r}, \vec{p})$ in eq.~(\ref{Eq-58}). In particular, we obtain 
\begin{equation}
e^{(d)}_{\text{F}} 
= \omega^{(d)} \left( d!~ N \right)^{1 / d} + e^{(d)}_{0} 
\label{Eq-61}
\end{equation}
in the $d$-dimensional TFA with the mean trap frequencies, $\omega^{(1)} \equiv \omega_{z}$, $\omega^{(2)} \equiv \omega_{c}$, and $\omega^{(3)} \equiv \omega_{c}^{2 / 3} \omega_{z}^{1 / 3}$, and zero-point energy shifts, $e^{(1)}_{0} \equiv \omega_{c}$, $e^{(2)}_{0} \equiv \omega_{z} / 2$, and $e^{(3)}_{0} \equiv 0$, according to eqs.~(\ref{Eq-51}), (\ref{Eq-52}), and (\ref{Eq-54}) or eq.~(\ref{Eq-B7}) in general. 

\begin{figure}[ht]
  \begin{center}
    \begin{tabular}{c}
       \includegraphics[scale=0.5,angle=-90]{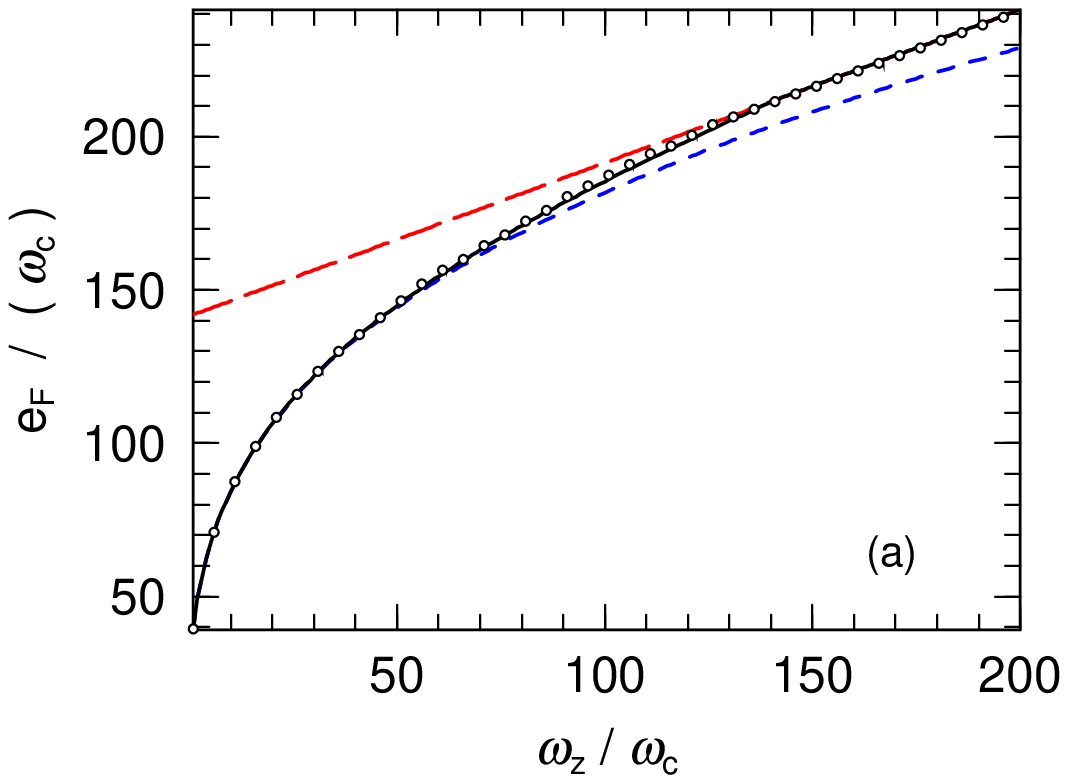} \\ 
       \includegraphics[scale=0.5,angle=-90]{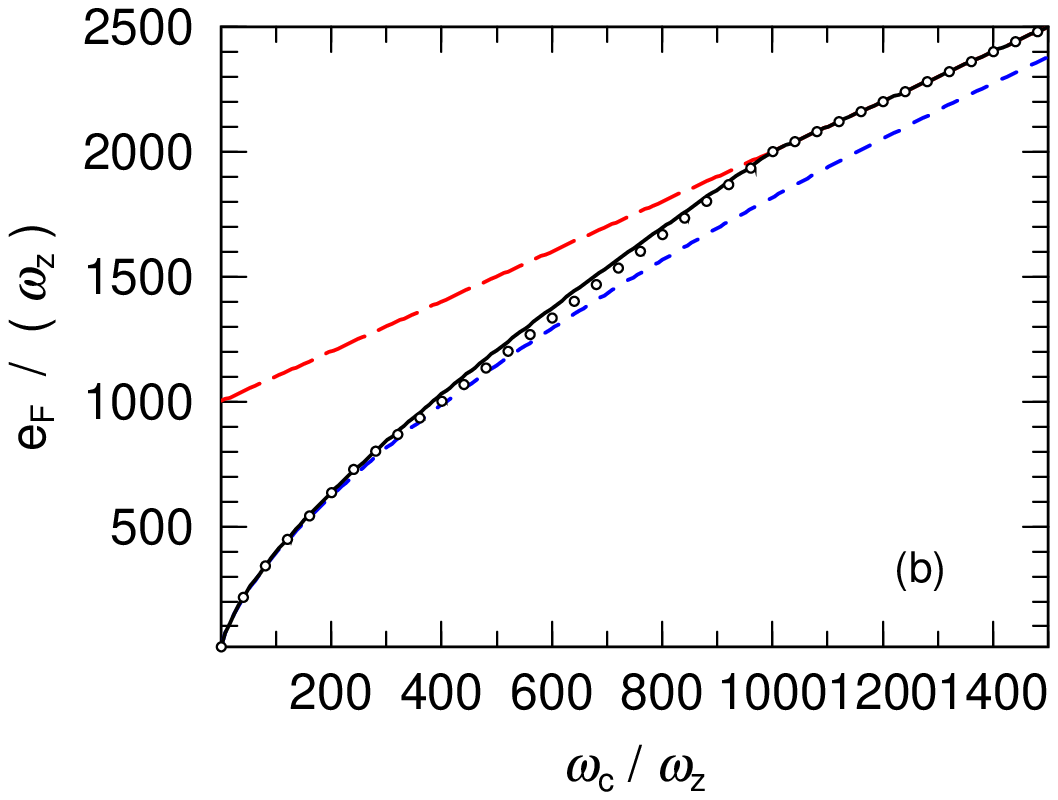} 
    \end{tabular}
  \end{center}
\caption{(Color online) 
The Fermi energies for the oblate deformed gases when $N = 10^{4}$ (a) and prolate deformed gases when $N = 10^{3}$ (b). The open circles represent the exact results. The solid lines represent the results in ETFA with $\varepsilon_{\text{c}} = 1.5 \omega_{z}$ (a) or $2 \omega_{c}$ (b). The short-dashed and long-dashed lines represent the results in 3DTFA and 2DTFA (a) or 1DTFA (b), respectively. 
}
\label{Fig-fe}
\end{figure}

In fig.~\ref{Fig-fe}, we show the Fermi energies for the oblate deformed gases when $N = 10^{4}$ (a) and prolate deformed gases when $N = 10^{3}$ (b). If the trap anisotropy is not so large, the exact values of the Fermi energies $e_{\text{F}}$ (denoted by the open circles) approximately agree with those in 3DTFA (denoted by the short-dashed lines). As the anisotropy increases, the Fermi energies $e_{\text{F}}$ exhibit the crossover behaviors from those in 3DTFA to those in 2DTFA (a) or 1DTFA (b). In fig.~\ref{Fig-fe}(a), the exact values agree with those in 2DTFA (denoted by the long-dashed line) when $e_{\text{F}} \le 1.5 \omega_{z}$ and $N \le 0.5 ( \omega_{z} / \omega_{c} )^{2}$. In fig.~\ref{Fig-fe}(b), the exact values agree with those in 1DTFA (denoted by the long-dashed line) when $e_{\text{F}} \le 2 \omega_{c}$ and $N \le \omega_{c} / \omega_{z}$. 

Furthermore, in fig.~\ref{Fig-fe}, we also plot the results in ETFA (denoted by the solid lines) including the QLD structures only below the cutoff energy, $\varepsilon_{\text{c}} = 1.5 \omega_{z}$ (a) or $\varepsilon_{\text{c}} = 2 \omega_{c}$ (b), for the lowest hierarchy, $q_{z} = 0$ (a) or $q_{c} = 0$ (b). These results roughly reproduce the crossover behaviors and are slightly different from the exact values in the middle range of the anisotropy owing to the upper hierarchies of the QLD structures. 

\begin{figure}[ht]
  \begin{center}
    \begin{tabular}{c}
       \includegraphics[scale=0.4,angle=-90]{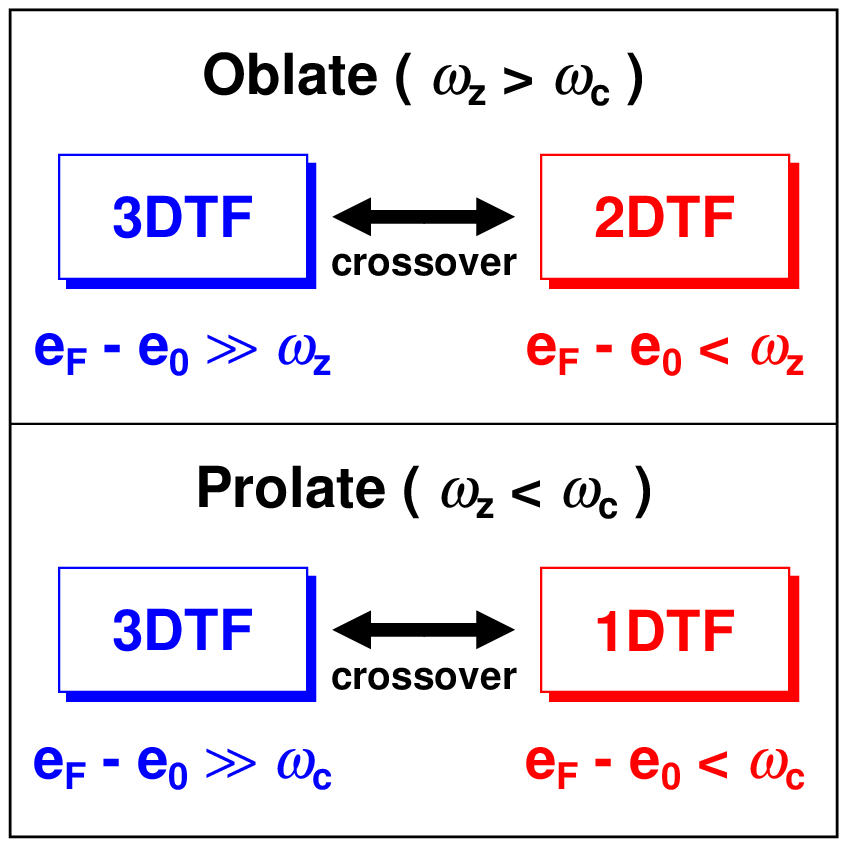} 
    \end{tabular}
  \end{center}
\caption{(Color online) 
The phase structures of the weakly-interacting Fermi gases in the highly-anisotropic traps. 
}
\label{Fig-pd}
\end{figure}

As shown in fig.~\ref{Fig-fe}, the QLD structures induce the crossover behaviors between the 3D and QLD gases as an important feature of the quantum gases in the highly-anisotropic traps. According to dimensional analysis, the phase structures of the weakly-interacting Fermi gases can be determined by $e_{\text{F}} / \omega_{z}$ for the oblate deformed gases and $e_{\text{F}} / \omega_{c}$ for the prolate deformed gases. Note that, in this case, the other energy scales, e.g., the other trap frequency and interaction energies, are very small and negligible in comparison to the Fermi energy $e_{\text{F}}$. We briefly describe the phase structures in fig.~\ref{Fig-pd}. 
\subsection{Determination of the parameters \label{Sec-III-B}}
Here we determine the parameters needed in the sum-rule-scaling method by using ETFA and the perturbation theory, where we expand these parameters by the coupling constant $g$ and take the zero-th and first order terms, e.g., $K_{c} \simeq K_{c 0} + K_{c 1}$, $U_{c} \simeq U_{c 0} + U_{c 1}$, and so on. 

According to the Wigner function $f(\vec{r}, \vec{p})$ in eq.~(\ref{Eq-55}), we obtain the zero-th order terms $U_{c 0}$ and $U_{z 0}$ of the potential energies in the $c$ and $z$-direction motions in eq.~(\ref{Eq-25-2}) as 
\begin{equation}
U_{c 0} 
= \sum_{n_{z} = 0}^{\infty} \frac{( e_{\text{F}} - e_{z} )^{3}}{6 \omega_{c}^{2}} \Theta{\left( e_{\text{F}} - e_{z} \right)} 
\label{Eq-62}
\end{equation}
and 
\begin{equation}
U_{z 0} 
= \sum_{n_{z} = 0}^{\infty} \frac{e_{z} ( e_{\text{F}} - e_{z} )^{2}}{4 \omega_{c}^{2}} \Theta{\left( e_{\text{F}} - e_{z} \right)} 
\label{Eq-63}
\end{equation}
for the oblate deformed gases and 
\begin{equation}
U_{c 0} 
= \sum_{n_{x} = 0}^{\infty} \sum_{n_{y} = 0}^{\infty} \frac{e_{c} ( e_{\text{F}} - e_{c} )}{2 \omega_{z}} \Theta{\left( e_{\text{F}} - e_{c} \right)} 
\label{Eq-64}
\end{equation}
and 
\begin{equation}
U_{z 0} 
= \sum_{n_{x} = 0}^{\infty} \sum_{n_{y} = 0}^{\infty} \frac{( e_{\text{F}} - e_{c} )^{2}}{4 \omega_{z}} \Theta{\left( e_{\text{F}} - e_{c} \right)} 
\label{Eq-65}
\end{equation}
for the prolate deformed gases, where we utilize $e_{z}$ and $e_{c} \equiv e_{x} + e_{y}$ in eq.~(\ref{Eq-42}) and integral formulas for the Hermite functions $H_{n}(x)$. 

The results in 3DTFA,  
\begin{equation}
U_{c 0}^{(3)} 
= 2 U_{z 0}^{(3)} 
= \frac{1}{4} \left( 6 \omega_{z} \omega_{c}^{2} N^{4} \right)^{1 / 3}, 
\label{Eq-66}
\end{equation}
can be reproduced by smoothing the residual sums in eqs.~(\ref{Eq-62})-(\ref{Eq-65}) as 
\begin{equation}
\sum_{n_{j} = 0}^{\infty} 
\to \int_{0}^{\infty} \frac{d{e_{j}}}{\omega_{j}} 
\label{Eq-66-0}
\end{equation}
for $j = x$, $y$, and $z$. 

The results in 2DTFA,  
\begin{equation}
U_{c 0}^{(2)} 
= \frac{\sqrt{2}}{3} \omega_{c} N^{3 / 2} 
\label{Eq-67-1}
\end{equation}
and 
\begin{equation}
U_{z 0}^{(2)} 
= \frac{\omega_{z}}{4} N, 
\label{Eq-67-2}
\end{equation}
can also be reproduced by taking $n_{z} = 0$ in the residual sums in eqs.~(\ref{Eq-62}) and (\ref{Eq-63}). 

The results in 1DTFA,  
\begin{equation}
U_{c 0}^{(1)} 
= \frac{\omega_{c}}{2} N 
\label{Eq-68-1}
\end{equation}
and 
\begin{equation}
U_{z 0}^{(1)} 
= \frac{\omega_{z}}{4} N^{2}, 
\label{Eq-68-2}
\end{equation}
can also be reproduced by taking $n_{x} = n_{y} = 0$ in the residual sums in eqs.~(\ref{Eq-64}) and (\ref{Eq-65}). 

\begin{figure}[ht]
  \begin{center}
    \begin{tabular}{cc}
       \includegraphics[scale=0.4,angle=-90]{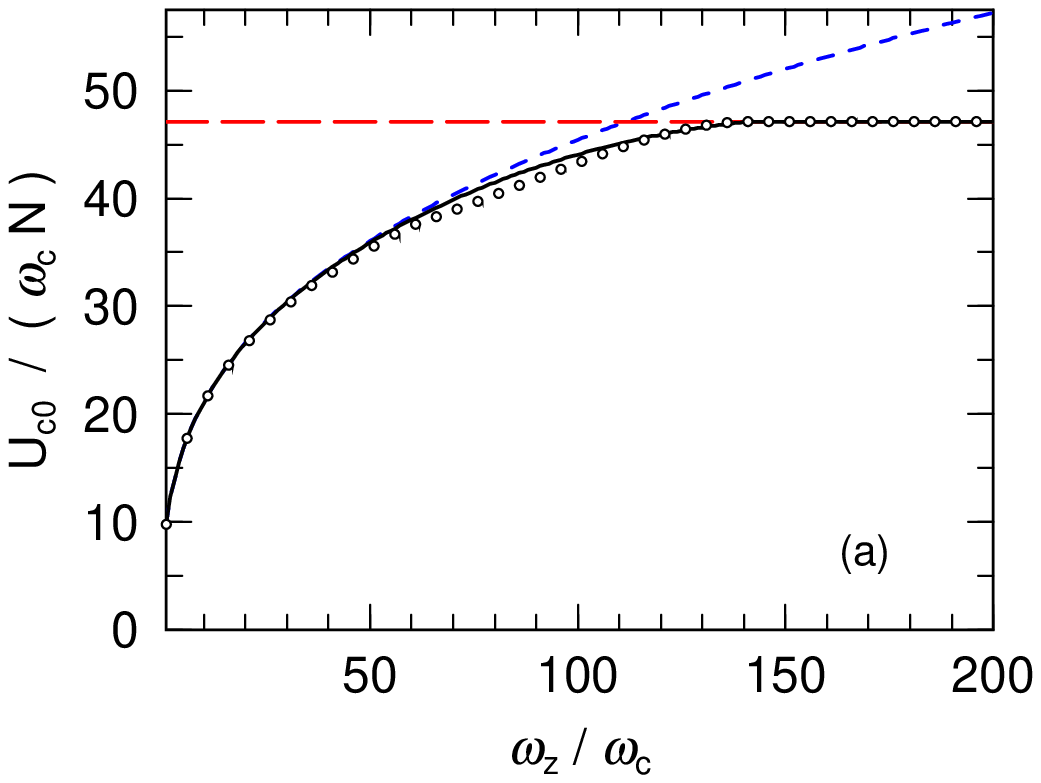} & 
       \includegraphics[scale=0.4,angle=-90]{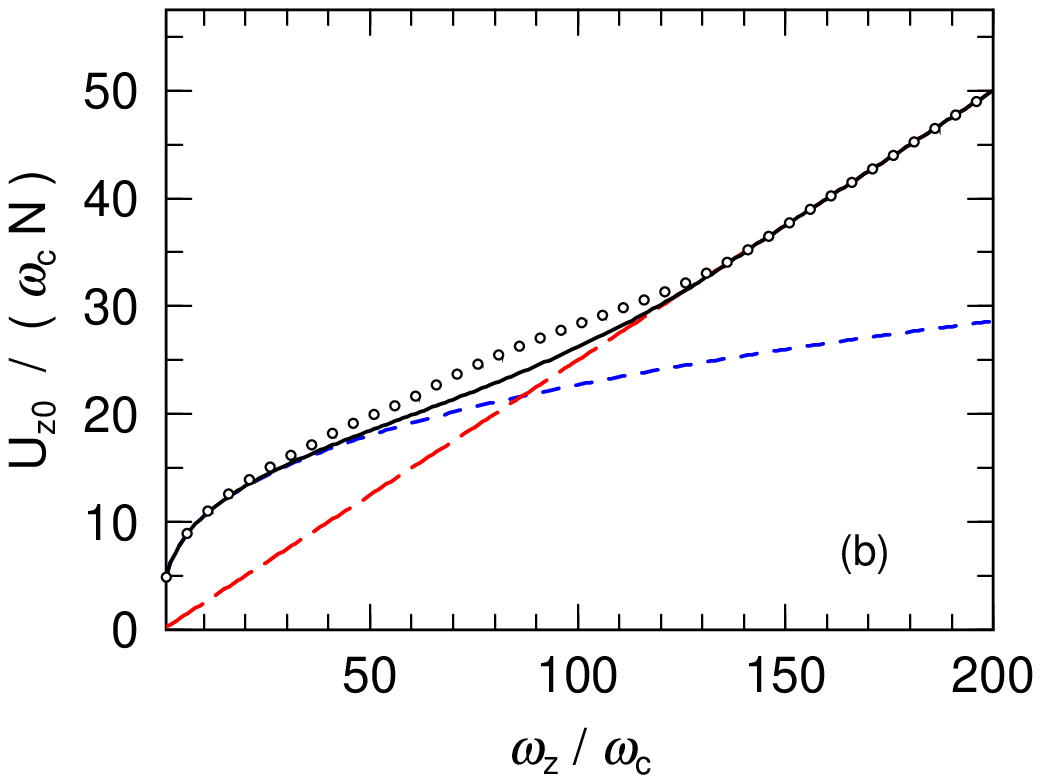} \\ 
       \includegraphics[scale=0.4,angle=-90]{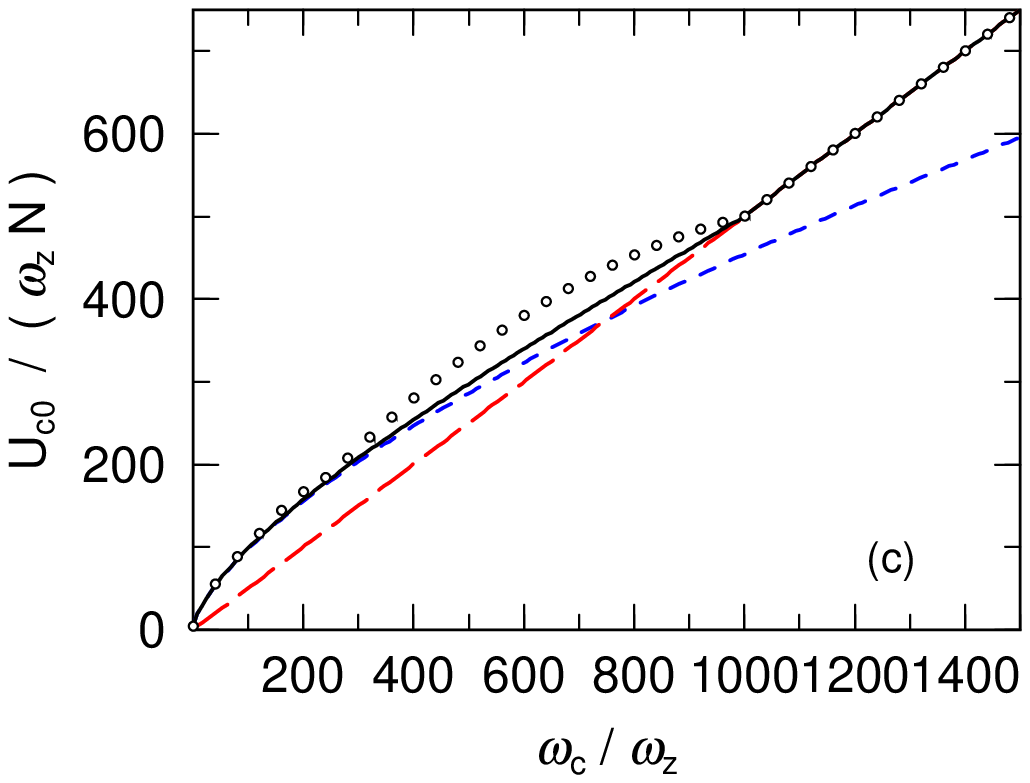} & 
       \includegraphics[scale=0.4,angle=-90]{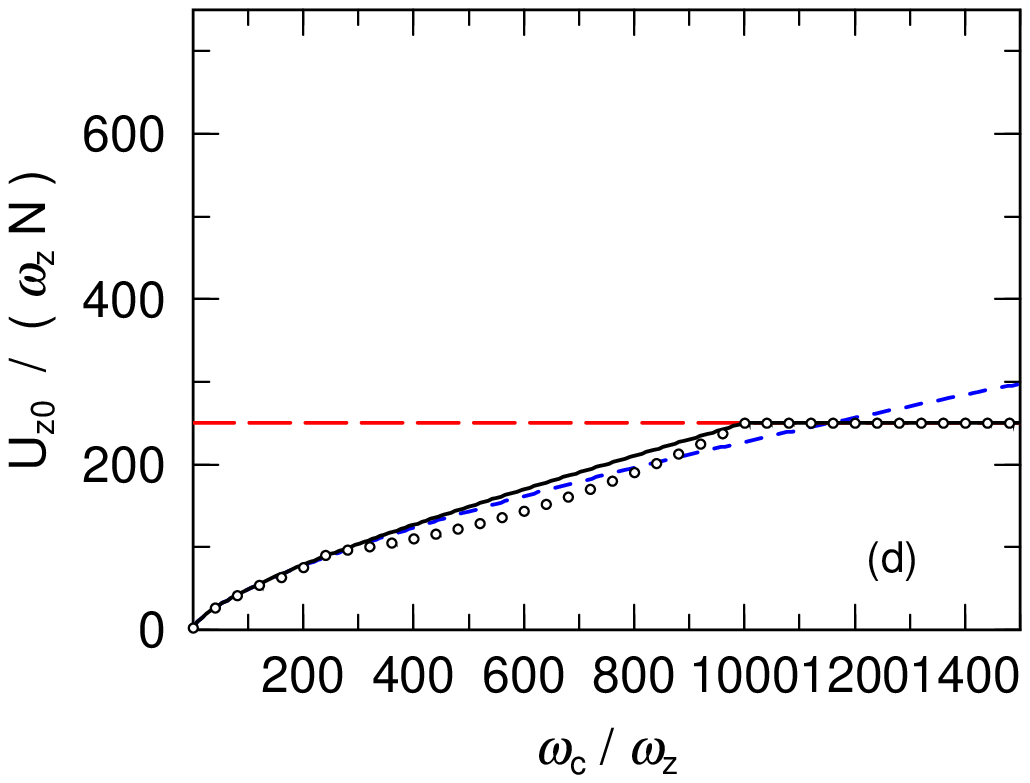} 
    \end{tabular}
  \end{center}
\caption{(Color online) 
The potential energies, $U_{c 0}$ [(a) (c)] and $U_{z 0}$ [(b) (d)], for the oblate deformed gases when $N = 10^{4}$ [(a) (b)] and prolate deformed gases when $N = 10^{3}$ [(c) (d)]. The open circles represent the exact results. The solid lines represent the results in ETFA with $\varepsilon_{\text{c}} = 1.5 \omega_{z}$ [(a) (b)] or $2 \omega_{c}$ [(c) (d)]. The short-dashed and long-dashed lines represent the results in 3DTFA and 2DTFA [(a) (b)] or 1DTFA [(c) (d)], respectively. 
}
\label{Fig-pe0}
\end{figure}

In Fig.~\ref{Fig-pe0}, we show the zero-th order terms $U_{c 0}$ and $U_{z 0}$ of the potential energies in eqs.~(\ref{Eq-62})-(\ref{Eq-65}) for the oblate deformed gases when $N = 10^{4}$ and prolate deformed gases when $N = 10^{3}$. The crossover behaviors and hierarchic structures also appear in these plots. 

The mass parameters $\mathcal{M}_{c}$ and $\mathcal{M}_{z}$ in eq.~(\ref{Eq-27}) are determined from the potential energies $U_{c}$ and $U_{z}$ owing to the definitions, 
\begin{equation}
\mathcal{M}_{j} 
= \frac{2}{\omega_{j}^{2}} U_{j} 
\label{Eq-69}
\end{equation}
for $j = c$ and $z$. 

The zero-th order terms $K_{c 0}$ and $K_{z 0}$ of the kinetic energies in eq.~(\ref{Eq-25-1}) are equal to those of the potential energies $U_{c 0}$ and $U_{z 0}$ according to the virial theorem in eqs.~(\ref{Eq-29}) and (\ref{Eq-30}), i.e., 
\begin{equation}
K_{j 0} 
= U_{j 0} 
\label{Eq-69-2}
\end{equation}
for $j = c$ and $z$. 

In addition, we can determine the first order terms $K_{c 1}$, $K_{z 1}$, $U_{c 1}$, and $U_{z 1}$ from the interaction energy $E_{\text{int}}$ as 
\begin{equation}
U_{c 1} 
= - K_{c 1} 
= \frac{1}{2} E_{\text{int}} 
\label{Eq-72-1}
\end{equation}
and 
\begin{equation}
U_{z 1} 
= - K_{z 1} 
= \frac{1}{4} E_{\text{int}} 
\label{Eq-72-2}
\end{equation}
because 
\begin{equation}
E_{\text{int}} 
= U_{c 1} - K_{c 1} 
= 2 ( U_{z 1} - K_{z 1} ) 
\label{Eq-71}
\end{equation}
according to the virial theorem in eqs.~(\ref{Eq-29}) and (\ref{Eq-30}) and 
\begin{equation}
U_{j 1} 
= - K_{j 1} 
\label{Eq-70}
\end{equation}
for $j = c$ and $z$ according to the first-order perturbation theory for the variable-separated non-perturbative state. 

Lastly we approximately determine the interaction energy $E_{\text{int}}$ as 
\begin{equation}
E_{\text{int}} 
\approx E_{\text{int}}^{(3)} 
\label{Eq-73-2}
\end{equation}
with that in 3DTFA, 
\begin{equation}
E_{\text{int}}^{(3)} 
= \beta^{(3)} g \sqrt{\omega_{z} \omega_{c}^{2} N^{3}}, 
\label{Eq-73-1-1}
\end{equation}
where $\beta^{(3)} \equiv 512 \sqrt{3} / (945 \pi^3) \approx 0.0302$. That is because the QLD structures have little influence on the interaction energy. In fact, the interaction energy in the crossover region varies between that in 3DTFA in eq.~(\ref{Eq-73-1-1}) and that in the $d$-dimensional TFA, 
\begin{equation}
E_{\text{int}}^{(d)} 
= \beta^{(d)} g \sqrt{\omega_{z} \omega_{c}^{2} N^{3}}, 
\label{Eq-73-1-2}
\end{equation}
for $d = 1$ or $2$, where $\beta^{(2)} \equiv 1 / (6 \pi^{3/2}) \approx 0.0299$ and $\beta^{(1)} \equiv 2 \sqrt{2} / (3 \pi^3) \approx 0.0304$, and they take nearly same values owing to $\beta^{(3)} \approx \beta^{(2)} \approx \beta^{(1)}$. 
\section{Breathing oscillations \label{Sec-IV}}
In this section, we show the calculational results with the parameters determined in the previous section. First we calculate the collective frequencies of the breathing oscillations in the sum-rule-scaling method in subsection~\ref{Sec-IV-A}. Second we demonstrate the time-evolution of the oscillations in the present framework in subsection~\ref{Sec-IV-B}. 
\subsection{Collective frequencies \label{Sec-IV-A}}
According to the eigen-equation, eq.~(\ref{Eq-32}), with eqs.~(\ref{Eq-27}) and (\ref{Eq-28}), we obtain the normal mode frequencies $\Omega_{c}$ and $\Omega_{z}$ of the breathing oscillations up to the first order of the interaction energy $E_{\text{int}}$ as 
\begin{equation}
\Omega_{j} 
= 2 \sqrt{\frac{K_{j} + U_{j}}{\mathcal{M}_{j}}} + \frac{\alpha_{j} E_{\text{int}}}{\sqrt{\mathcal{M}_{j} ( K_{j} + U_{j} )}} + O{\left( E_{\text{int}}^{2} \right)} 
\label{Eq-74}
\end{equation}
for $j = c$ and $z$, where $\alpha_{c} \equiv 1$ and $\alpha_{z} \equiv 1 / 4$. Furthermore these solutions can also be expanded by the coupling constant $g$ as $\Omega_{j} \simeq \Omega_{j 0} + \Omega_{j 1}$, where 
\begin{equation}
\Omega_{j 0} 
= 2 \sqrt{\frac{K_{j 0} + U_{j 0}}{\mathcal{M}_{j 0}}}, 
\label{Eq-74-1}
\end{equation}
and 
\begin{equation}
\Omega_{j 1} 
= \frac{\Omega_{j 0}}{2} \left( \frac{K_{j 1} + U_{j 1}}{K_{j 0} + U_{j 0}} - \frac{\mathcal{M}_{j 1}}{\mathcal{M}_{j 0}} + \frac{\alpha_{j} E_{\text{int}}}{K_{j 0} + U_{j 0}} \right). 
\label{Eq-74-2}
\end{equation}

The zero-th order terms in eq.~(\ref{Eq-74-1}) correspond to those of the noninteracting gases. In fact, by substituting eqs.~(\ref{Eq-69}) and (\ref{Eq-69-2}) into eq.~(\ref{Eq-74-1}), we obtain 
\begin{equation}
\Omega_{j 0} 
= 2 \omega_{j}
\label{Eq-75}
\end{equation}
for $j = c$ and $z$. The results in eq.~(\ref{Eq-75}) depend only on the trap frequencies and are independent of the QLD structures. 

The first order terms $\Omega_{c 1}$ and $\Omega_{z 1}$ in eq.~(\ref{Eq-74-2}) can also be obtained as 
\begin{equation}
\Omega_{c 1} 
= 0 
\label{Eq-76-1}
\end{equation}
and 
\begin{equation}
\Omega_{z 1} 
= - \frac{\omega_{z}}{8} \frac{E_{\text{int}}}{U_{z 0}} 
\label{Eq-76-2}
\end{equation}
by substituting eqs.~(\ref{Eq-69})-(\ref{Eq-72-2}) into eq.~(\ref{Eq-74-2}). The results in eqs.~(\ref{Eq-76-1}) and (\ref{Eq-76-2}) reveal that the interaction and QLD structures affect the longitudinal normal mode $\Omega_{z}$, but not the transverse normal mode $\Omega_{c}$. Note that the vanishment of the first order term $\Omega_{c 1}$ in eq.~(\ref{Eq-76-1}) must be a universal property of the transverse mode $\Omega_{c}$ owing to the structure of the off-diagonal elements in eq.~(\ref{Eq-28}). 

According to eq.~(\ref{Eq-76-2}), the QLD structures contribute the longitudinal normal mode $\Omega_{z}$ principally through the potential energy $U_{z 0}$ because eq.~(\ref{Eq-76-2}) contains only the two parameters, $U_{z 0}$ and $E_{\text{int}}$, and the QLD structures have little influence on $E_{\text{int}}$ as shown in eq.~(\ref{Eq-73-2}). As a result, the longitudinal normal mode $\Omega_{z}$ exhibits the crossover behaviors associated with those of the potential energy $U_{z 0}$ shown in fig.~\ref{Fig-pe0}. 

According to eqs.~(\ref{Eq-66})-(\ref{Eq-68-2}), the first order term $\Omega_{z 1}$ of the longitudinal normal mode in eq.~(\ref{Eq-76-2}) becomes 
\begin{equation}
\Omega_{z 1}^{(3)} 
= - \frac{1}{6^{1 / 3}} \left( \frac{\omega_{z}}{\omega_{c}} \right)^{2 / 3} \frac{E_{\text{int}}}{N^{4 / 3}} 
\label{Eq-77}
\end{equation}
in the 3DTFA limit, 
\begin{equation}
\Omega_{z 1}^{(2)} 
= - \frac{1}{2} \frac{E_{\text{int}}}{N} 
\label{Eq-78}
\end{equation}
in the 2DTFA limit, and 
\begin{equation}
\Omega_{z 1}^{(1)} 
= - \frac{1}{2} \frac{E_{\text{int}}}{N^{2}} 
\label{Eq-79}
\end{equation}
in the 1DTFA limit. 

\begin{figure}[ht]
  \begin{center}
    \begin{tabular}{cc}
       \includegraphics[scale=0.4,angle=-90]{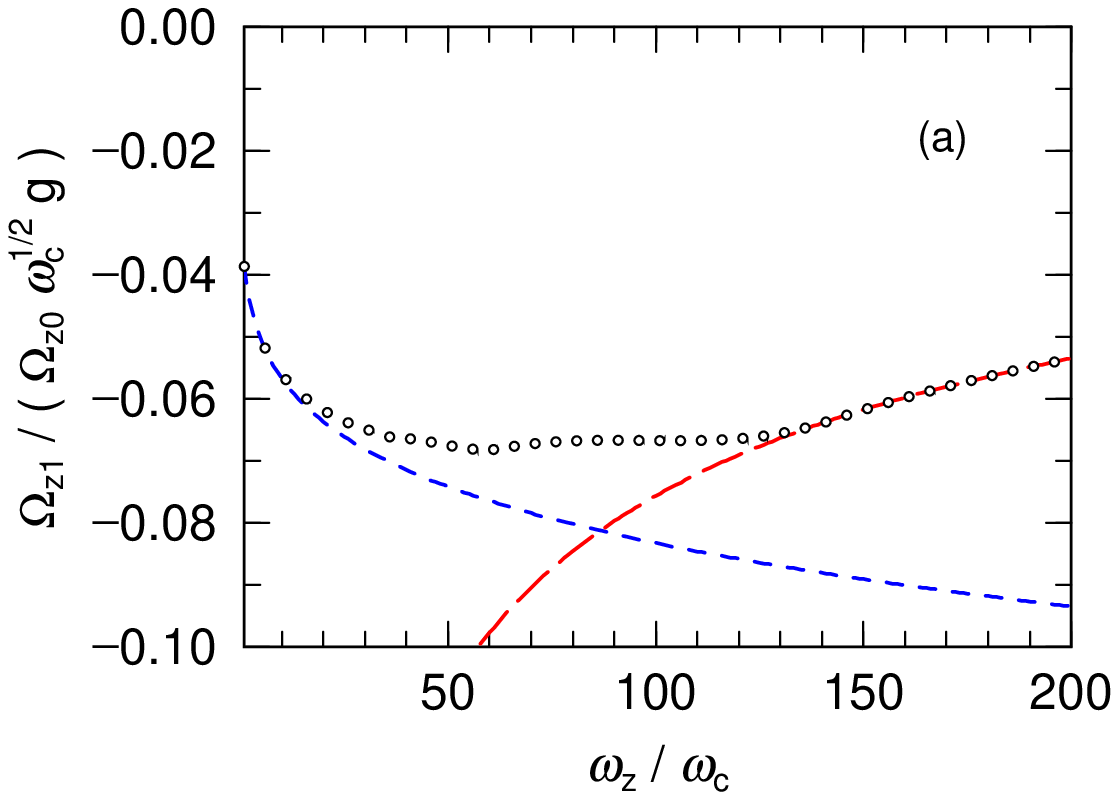} & 
       \includegraphics[scale=0.4,angle=-90]{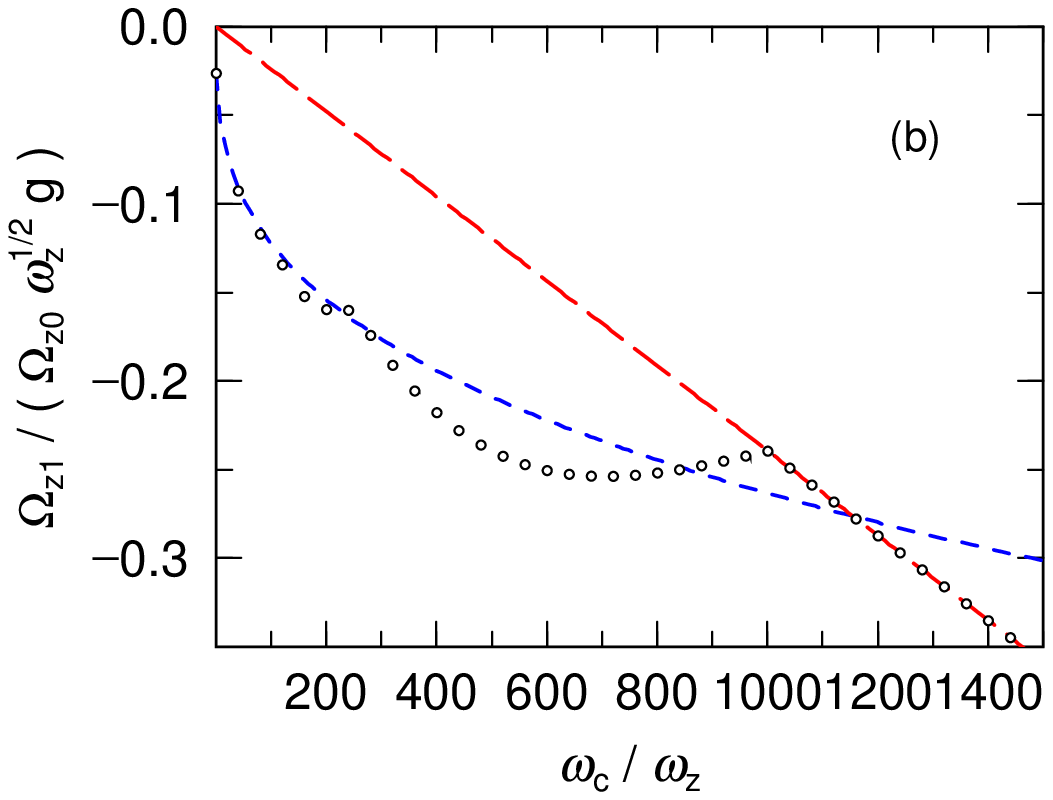} 
    \end{tabular}
  \end{center}
\caption{(Color online) 
The first order term of the longitudinal normal mode, $\Omega_{z 1}$, for the oblate deformed gases when $N = 10^{4}$ (a) and prolate deformed gases when $N = 10^{3}$ (b). The open circles represent the exact results. The short-dashed and long-dashed lines represent the results in 3DTFA and 2DTFA (a) or 1DTFA (b), respectively. 
}
\label{Fig-oz1}
\end{figure}

In fig.~\ref{Fig-oz1}, we show the first order term $\Omega_{z 1}$ of the longitudinal normal mode in eq.~(\ref{Eq-76-2}) with the parameters in eqs.~(\ref{Eq-63}), (\ref{Eq-65}), and (\ref{Eq-73-2}) for the oblate deformed gases when $N = 10^{4}$ (a) and prolate deformed gases when $N = 10^{3}$ (b). The situations are same as those in figs.~\ref{Fig-fe} and \ref{Fig-pe0}. The exact results (denoted by the open circles) exhibit the crossover behaviors between the results in 3DTFA (denoted by the short-dashed lines) and 2DTFA (a) or 1DTFA (b) (denoted by the long-dashed lines). 

\begin{figure}[ht]
  \begin{center}
    \begin{tabular}{cc}
       \includegraphics[scale=0.4,angle=-90]{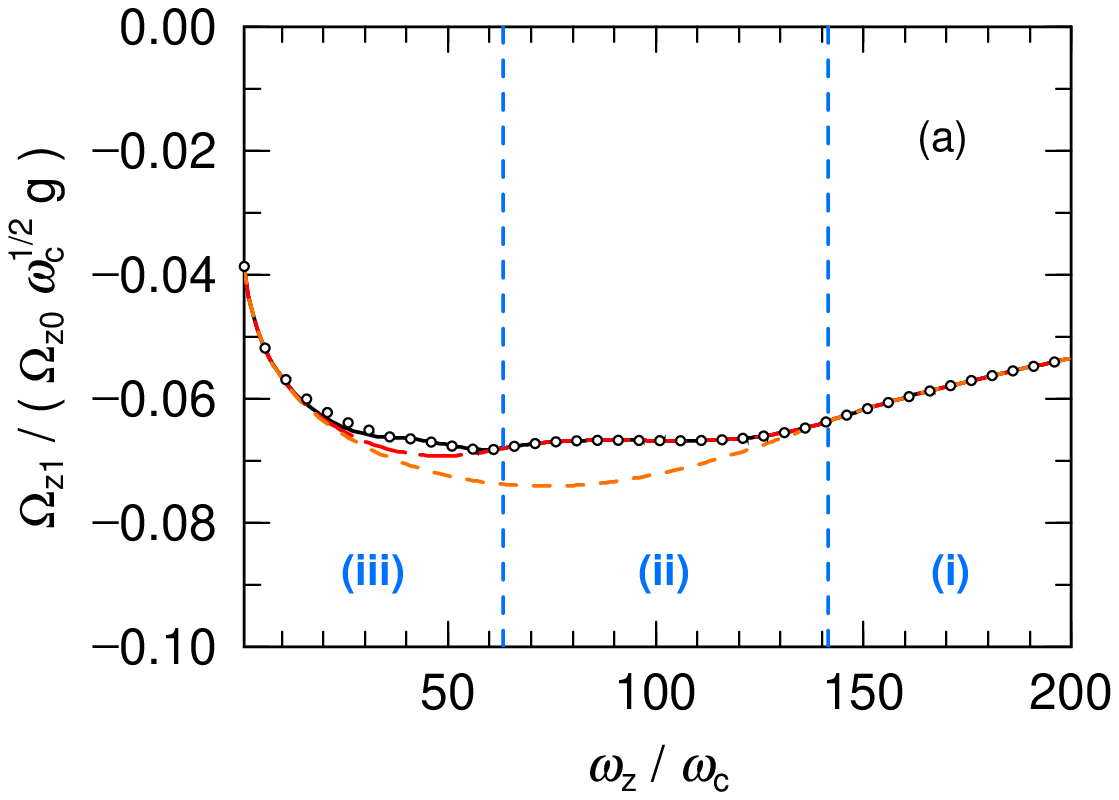} & 
       \includegraphics[scale=0.4,angle=-90]{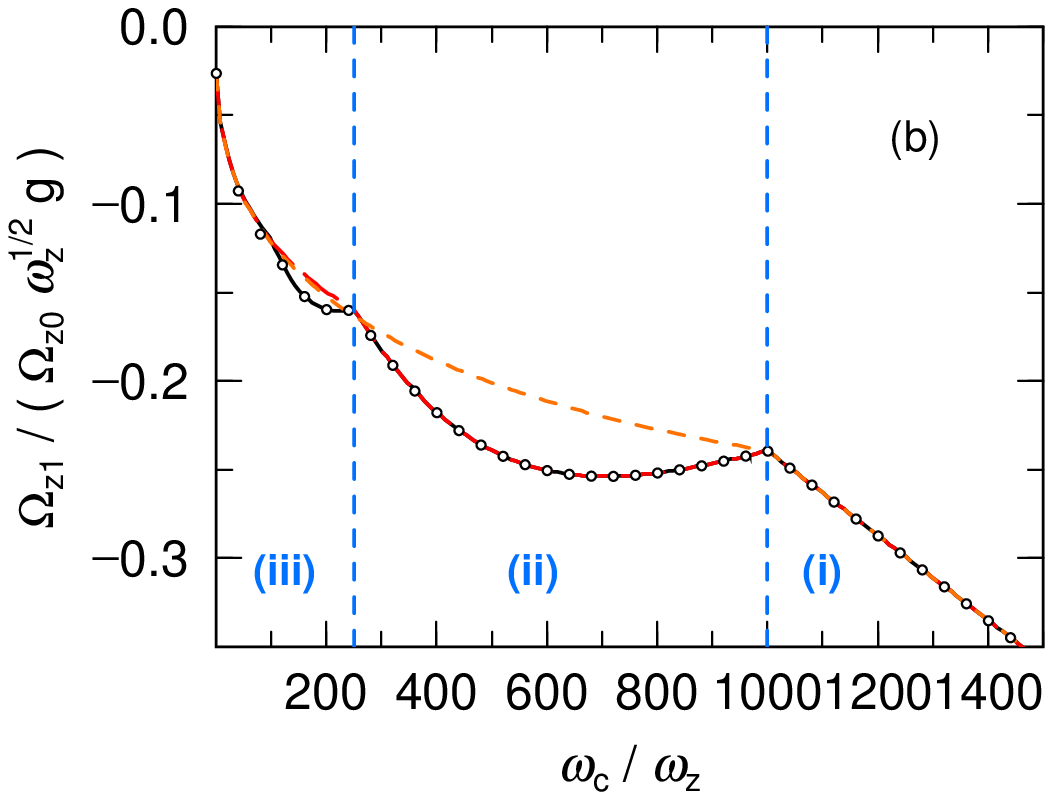} 
    \end{tabular}
  \end{center}
\caption{(Color online) 
Same as fig.~\ref{Fig-oz1}, but the short-dashed, long-dashed, and solid lines represent the results in ETFA with $\varepsilon_{\text{c}} = 1.5 \omega_{z}$, $2.5 \omega_{z}$, and $3.5 \omega_{z}$ (a) and $2 \omega_{c}$, $3 \omega_{c}$, and $4 \omega_{c}$ (b), respectively. 
}
\label{Fig-oz1-etfa}
\end{figure}

In fig.~\ref{Fig-oz1-etfa}, we show comparison between the exact results in fig.~\ref{Fig-oz1} (denoted by the open circles) and the results in ETFA with $\varepsilon_{\text{c}} = 1.5 \omega_{z}$, $2.5 \omega_{z}$, and $3.5 \omega_{z}$ (a) and $2 \omega_{c}$, $3 \omega_{c}$, and $4 \omega_{c}$ (b) (denoted by the short-dashed, long-dashed, and solid lines, respectively). There clearly appear three ranges reflecting the hierarchic structures. In fig.~\ref{Fig-oz1-etfa}(a), we present (i) $\omega_{z} > \sqrt{2 N} \omega_{c}$, (ii) $\sqrt{2 N} \omega_{c} > \omega_{z} > \sqrt{0.4 N} \omega_{c}$, and (iii) $\sqrt{0.4 N} \omega_{c} > \omega_{z}$. In fig.~\ref{Fig-oz1-etfa}(b), we present (i) $\omega_{c} > N \omega_{z}$, (ii) $N \omega_{z} > \omega_{c} > 0.25 N \omega_{z}$, and (iii) $0.25 N \omega_{z} > \omega_{c}$. 

In range (i), the exact results agree with those in 2DTFA (a) or 1DTFA (b), and all results in ETFA reproduce the exact values. In other words, only the lowest hierarchy in the QLD structures contributes to the results in this range. 

In range (ii), the lowest and second-lowest hierarchies contribute to the exact results. In fact, the results in ETFA only with $\varepsilon_{\text{c}} \ge 2.5 \omega_{z}$ (a) or $3 \omega_{c}$ (b) reproduce the exact values. The difference between the results in the exact calculation and ETFA with $\varepsilon_{\text{c}} = 1.5 \omega_{z}$ (a) or $2 \omega_{c}$ (b) indicates the contribution of the second-lowest hierarchy. 

In range (iii), the lowest, second-lowest, and third-lowest hierarchies principally contribute to the exact results. Thus the results in ETFA only with $\varepsilon_{\text{c}} \ge 3.5 \omega_{z}$ (a) or $5 \omega_{c}$ (b) can reproduce the exact values. There appear additional detailed structures owing to the upper hierarchies; However it is difficult to see them in fig.~\ref{Fig-oz1-etfa} because the difference between the results in the exact calculation and ETFA is visually small. 

Finally we comment on the particle number dependence on the above results. If the particle number varies, the qualitative behaviors are kept, and two kinds of quantitative change appear. One is the change of the results in the TFA limit, $\Omega^{(3)}_{z 1} \propto N^{1 / 6}$, $\Omega^{(2)}_{z 1} \propto N^{1 / 2}$, and $\Omega^{(1)}_{z 1} \propto N^{- 1 / 2}$, as shown in eqs.~(\ref{Eq-77})-(\ref{Eq-79}). The other is the change of the critical anisotropies dividing range (i) and range (ii), which are proportional to $N^{1 / 2}$ for the oblate deformed gases and $N$ for the prolate deformed gases. 
\subsection{Time-evolution \label{Sec-IV-B}}
In the sum-rule-scaling method, the collective coordinates $\lambda_{c}(t)$ and $\lambda_{z}(t)$ determine the breathing oscillations as described in eq.~(\ref{Eq-13}). In order to observe the collective coordinates in actual experiments, we can utilize the measurement of the projected mean square radii $s_{c}(t)$ and $s_{z}(t)$ defined in eq.~(\ref{Eq-11-0}) because they are proportional to $\lambda_{c}(t)$ and $\lambda_{z}(t)$, respectively. 

Here we demonstrate the time-evolution of $s_{c}(t)$ ($\propto \lambda_{c}(t)$) and $s_{z}(t)$ ($\propto \lambda_{z}(t)$) by solving the classical dynamics denoted by eqs.~(\ref{Eq-20}) and (\ref{Eq-26}). Note that the collective coordinates $\lambda_{c}(t)$ and $\lambda_{z}(t)$ are defined as the variational parameters in the sum-rule-scaling method and contain the quantum fluctuations as mean-values in TDHFA. 

Up to the first order of the the interaction energy $E_{\text{int}}$, the equations of motion can be described as 
\begin{equation}
\frac{d^{2}{\lambda_{c}}}{d{t}^{2}} 
= - \Omega_{c}^{2} \left( \lambda_{c} + \gamma_{c} \lambda_{z} \right) 
\label{Eq-80-1}
\end{equation}
and 
\begin{equation}
\frac{d^{2}{\lambda_{z}}}{d{t}^{2}} 
= - \Omega_{z}^{2} \left( \lambda_{z} + \gamma_{z} \lambda_{c} \right) 
\label{Eq-80-2}
\end{equation}
with 
\begin{equation}
\gamma_{j} 
\equiv \frac{\omega_{j}^{2}}{\Omega_{j}^{2}} \frac{E_{\text{int}}}{U_{j}} 
\label{Eq-81}
\end{equation}
for $j = c$ and $z$. 

\begin{figure}[ht]
  \begin{center}
    \begin{tabular}{cc}
       \includegraphics[scale=0.4,angle=-90]{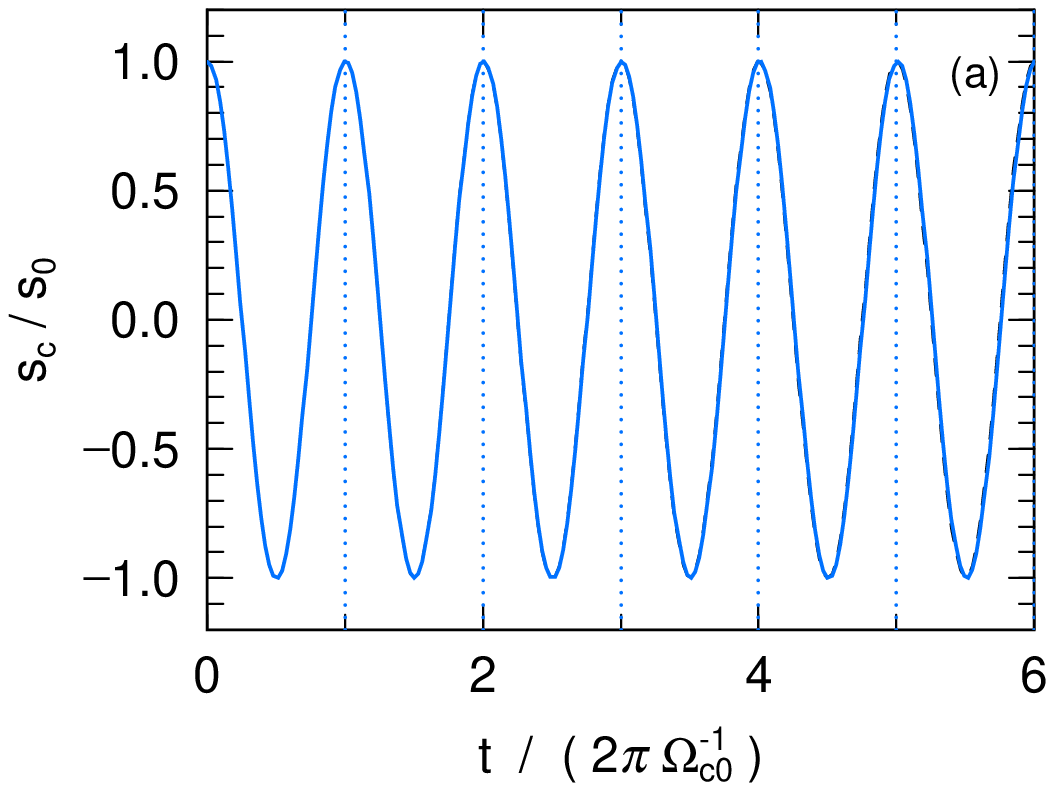} &
       \includegraphics[scale=0.4,angle=-90]{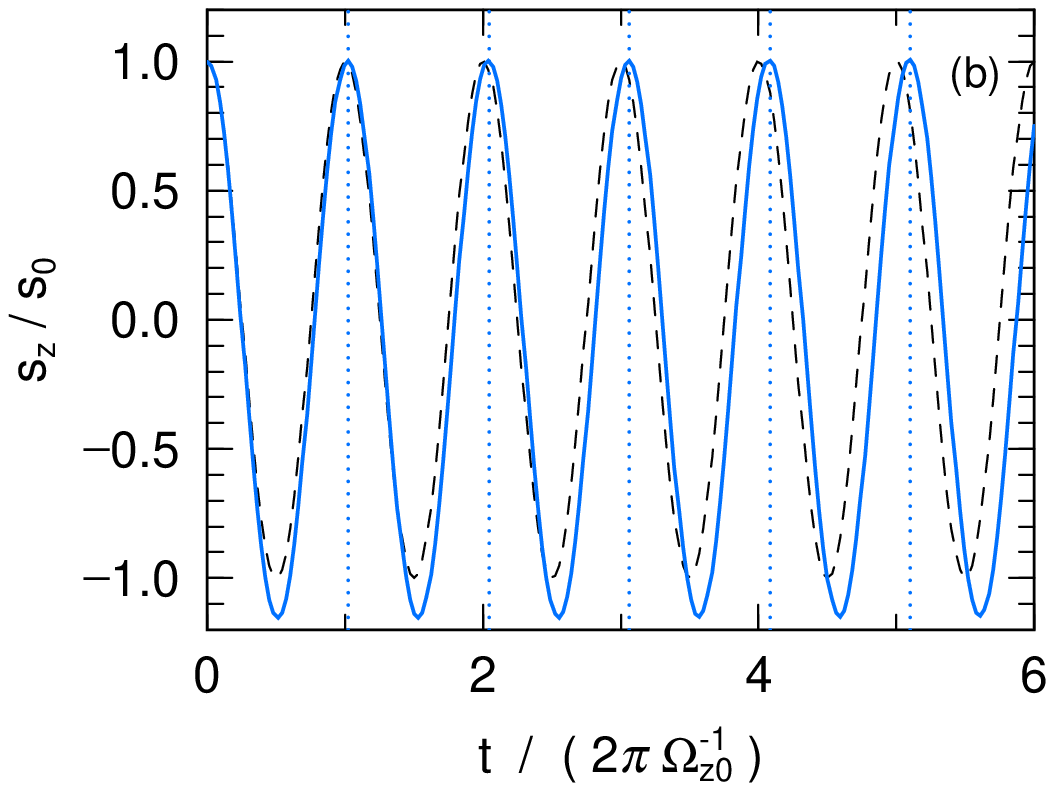} \\ 
       \includegraphics[scale=0.4,angle=-90]{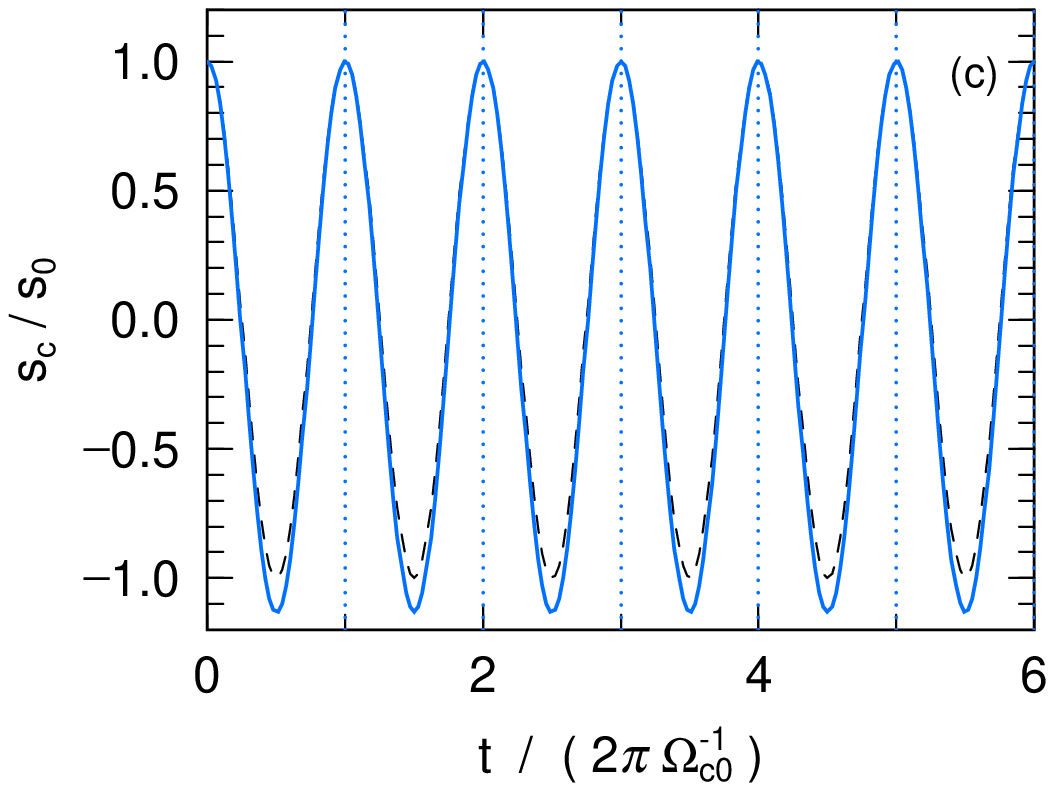} &
       \includegraphics[scale=0.4,angle=-90]{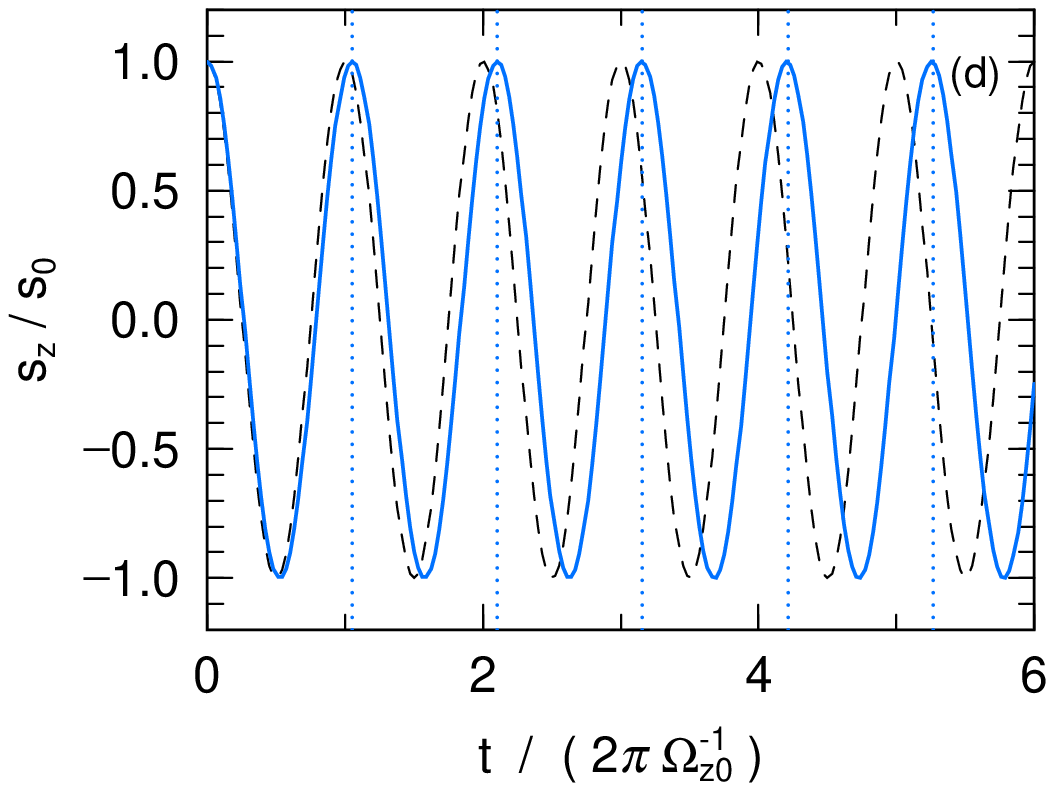} 
    \end{tabular}
  \end{center}
\caption{(Color online) 
The time evolution of $s_{c}(t)$ [(a) (c)] and $s_{z}(t)$ [(b) (d)] of the oblate deformed gases when $\omega_{z} / \omega_{c} = 80$ and $N = 10^{4}$ [(a) (b)] and prolate deformed gases when $\omega_{c} / \omega_{z} = 600$ and $N = 10^{3}$ [(c) (d)]. The short-dashed lines represent the results for $g = 0$. The solid lines and dotted separations represent the results and normal mode periods, respectively, for $g = 0.3 \omega_{c}^{- 1 / 2}$ [(a) (b)] and $0.2 \omega_{z}^{- 1 / 2}$ [(c) (d)]. 
}
\label{Fig-sim}
\end{figure}

In Fig.~\ref{Fig-sim}, we show the time evolution of $s_{c}(t)$ [(a) (c)] and $s_{z}(t)$ [(b) (d)] (denoted by the solid lines) for the oblate deformed gases when $g = 0.3 \omega_{c}^{- 1 / 2}$, $\omega_{z} / \omega_{c} = 80$, and $N = 10^{4}$ [(a) (b)] and prolate deformed gases when $g = 0.2 \omega_{z}^{- 1 / 2}$, $\omega_{c} / \omega_{z} = 600$, and $N = 10^{3}$ [(c) (d)]. Here we give an initial condition as $s_{c}(t = 0) = s_{z}(t = 0) = s_{0}$ and $\dot{s}_{c}(t = 0) = \dot{s}_{z}(t = 0) = 0$. This condition corresponds to the typical experimental situation, where the trap frequencies are suddenly changed at $t = 0$. For reference, we also plot the results of the non-interacting gases (denoted by the short-dashed lines) and periods of the normal modes (denoted by the dotted separations). 

As shown in fig.~\ref{Fig-sim}, the interaction principally induces two kinds of influence. One is the frequency decrement in the longitudinal oscillations shown in fig.~\ref{Fig-sim}(b) and (d). The other is the shift of the center position of the faster oscillations shown in fig.~\ref{Fig-sim}(b) and (c). 

The frequency decrement in the longitudinal oscillations shown in fig.~\ref{Fig-sim}(b) and (d) can be explained from the normal mode frequencies. According to eqs.~(\ref{Eq-76-1}) and (\ref{Eq-76-2}), the interaction keeps the normal mode frequencies in the transverse oscillations and reduces (or increases) those in the longitudinal oscillations when $g > 0$ (or $< 0$). As shown in fig.~\ref{Fig-sim}, the frequencies of the plotted oscillations agree with the corresponding normal mode frequencies owing to eqs.~(\ref{Eq-80-1}) and (\ref{Eq-80-2}). As a result, the frequency decrement occurs only in the longitudinal oscillations. 

The center shift in the faster oscillations shown in fig.~\ref{Fig-sim}(b) and (c) is induced by the mixing of the faster and slower oscillations. The mixing effect of the faster oscillations on the slower oscillations approximately vanishes because of the coarse graining in the long time scales corresponding to the periods of the slower oscillations~\cite{footnote-1}. On the other hand, the mixing effect of the slower oscillations on the faster oscillations remains as the center shift, which is nearly constant in the small time scales and exhibit the beat structure in the long time scales. In addition, the center shift can also be seen in eqs.~(\ref{Eq-80-1}) and (\ref{Eq-80-2}). In fact, when $\lambda_{z}$ in eq.~(\ref{Eq-80-1}) (or $\lambda_{c}$ in eq.~(\ref{Eq-80-2})) is nearly constant in time as the slower oscillations, the mixing term $\gamma_{c} \lambda_{z}$ in eq.~(\ref{Eq-80-1}) (or $\gamma_{z} \lambda_{c}$ in eq.~(\ref{Eq-80-2})) gives the center shift. Note that the mixing effect principally appears in the center shift of the faster oscillations because the mixing term does not affect the normal mode frequencies in the first order terms. 

Lastly we point out that the faster oscillations can exhibit strong collectivity in the highly-anisotropic limits, where the fermions take the same quantum number in the narrower direction motion, i.e., the lowest hierarchy in the QLD structures, and their collective single-particle wave-functions principally determine the faster oscillations in analogy with the collective oscillations of the BECs. 
\section{Summary and outlook \label{Sec-V}}
In the present paper, we study the breathing oscillations of the weakly-interacting degenerate Fermi gases in the highly-anisotropic traps. If the traps are not highly anisotropic, the gases behave as the 3D gases; Otherwise the gases exhibit the QLD properties derived from the QLD structures. We focus on the effects of the QLD structures on the breathing oscillations of the symmetric fermions in the crossover range between the 3D and QLD gases. 

The breathing oscillations can be formulated in TDHFA and the sum-rule-scaling method as described in section~\ref{Sec-II}. The minimal collective oscillations can reflect the detailed properties of the ground and excited states of the trapped quantum gases. The QLD structures contribute to the oscillations through the ground state properties in the sum-rule-scaling method. 

The ground state properties with the QLD structures can be described in ETFA as explained in section~\ref{Sec-III}. The highly-anisotropic deformed gases exhibit the crossover behaviors with the hierarchic structures. The hierarchies can be reproduced by ETFA with the finite cutoff energy $\varepsilon_{\text{c}}$ and clearly shown in the perturbation theory for the weakly-interacting gases. In addition, we there obtain the parameters needed in the sum-rule-scaling method. 

The collective frequencies are calculated in the sum-rule-scaling method and perturbation theory in section~\ref{Sec-IV}. As a result, we reveal that the effects of the interaction and QLD structures simultaneously appear only in the longitudinal modes with the hierarchies and can not be seen in the transverse modes in the first-order perturbation. Finally we also demonstrate the time-evolution of the oscillations in the present framework, where we find out two kinds of the influence of the interaction: the frequency decrement (or increment) in the longitudinal oscillations when $g > 0$ (or $< 0$) and the center shift in the faster oscillations. 

The results in the present paper offer theoretical predictions for the weakly-interacting gases in actual experiments. If the inter-particle interaction and correlation are very strong, the present approach is not applicable, and the oscillation properties may be different from those in the present results. In addition, if the oscillations include more than one strong modes, e.g., near the phase-separation or collapse, the sum-rule-scaling method in the present framework should not be applied. In those cases, we should perform more detailed investigation in the full-microscopic RPA or other direct time-evolution approaches compatible with the highly-anisotropic systems~\cite{TM-01, TM-02}. 

Asymmetric two-component gases in the highly-anisotropic traps must be another interesting topic. In this case, the breathing oscillations implicitly include not only the in-phase oscillations but also the out-of-phase oscillations and exhibit the coupled oscillations of the in-phase and out-of-phase modes~\cite{2007-01-TMTN}. We should study the ground state properties and collective excitations in another paper. 
\acknowledgments
The authors would like to thank Prof. Toru Suzuki for useful discussions. This work was supported by KAKENHI (21540412 and 22540414). 
\appendix
\section{The scaling and sum-rule methods \label{App-A}}
In the scaling method, the density matrix $\varrho$ in the collective oscillation is derived from the density matrix $\varrho_{0}$ in equilibrium as 
\begin{equation}
\varrho{\big( \lambda, \dot{\lambda} \big)} 
= U{\big( \lambda, \dot{\lambda} \big)} \varrho_{0} U^{\dagger}{\big( \lambda, \dot{\lambda} \big)} 
\label{Eq-A1}
\end{equation}
with the excitation unitary operator 
\begin{equation}
U{\big( \lambda, \dot{\lambda} \big)} 
\equiv e^{i \dot{\lambda} \mathcal{O}} e^{\lambda \left[ H, \mathcal{O} \right]}, 
\label{Eq-A2}
\end{equation}
where $\lambda(t)$ and $\dot{\lambda}(t)$ indicate the time-dependent amplitude of the collective oscillation and its time derivative, respectively, and the Hermite operator $\mathcal{O}$ ($= \mathcal{O}^{\dagger}$) determines the property of the oscillation. Here the density matrices $\varrho$ and $\varrho_{0}$ are normalized, $\text{tr}{\left( \varrho \right)} = \text{tr}{\left( \varrho_{0} \right)} = 1$, and produce the expectation values as $\big< A \big> \equiv \text{tr}{\left( \varrho A \right)}$ and $\big< A \big>_{0} \equiv \text{tr}{\left( \varrho_{0} A \right)}$ for an arbitrary operator $A$. In addition, we here take $\big< \mathcal{O} \big>_{0} \equiv 0$ and $\big< H \big>_{0} \equiv 0$ in order to simplify the following description without loss of generality. 

The time-dependent variational principle, 
\begin{equation}
\delta \int d{t}~ \mathcal{L}{\big[ \lambda, \dot{\lambda} \big]} 
= 0 
\label{Eq-A3}
\end{equation}
with the Lagrangian 
\begin{equation}
\mathcal{L}{\big[ \lambda, \dot{\lambda} \big]} 
\equiv \left< i \frac{d}{d{t}} - H \right>, 
\label{Eq-A3-2}
\end{equation}
gives the Euler-Lagrange equation for the dynamical variable $\lambda(t)$. By substituting eq.~(\ref{Eq-A1}) into eq.~(\ref{Eq-A3-2}), we obtain 
\begin{eqnarray}
\mathcal{L} 
&\simeq& \frac{1}{2} \left< \big[ \big[ \mathcal{O}, H \big], \mathcal{O} \big] \right>_{0} \dot{\lambda}^{2} 
\nonumber \\ 
&-& \frac{1}{2} \left< \big[ \big[ \mathcal{O}, H \big], \big[ H, \big[ H, \mathcal{O} \big] \big] \big] \right>_{0} \lambda^{2}
\label{Eq-A4}
\end{eqnarray}
up to the second order of $\lambda(t)$. 

If the density matrices $\varrho_{0}$ and $\varrho$ are pure states, i.e., $\varrho_{0} = \big| \Psi_{0} \big> \big< \Psi_{0} \big|$, the coefficients in eq.~(\ref{Eq-A4}) become 
\begin{equation}
\frac{1}{2} \left< \big[ \big[ \mathcal{O}, H \big], \mathcal{O} \big] \right>_{0} 
= M_{1} 
\label{Eq-A5-1}
\end{equation}
and 
\begin{equation}
\frac{1}{2} \left< \big[ \big[ \mathcal{O}, H \big], \big[ H, \big[ H, \mathcal{O} \big] \big] \big] \right>_{0} 
= M_{3} 
\label{Eq-A5-2}
\end{equation}
with the $n$-th energy moment 
\begin{equation}
M_{n} 
\equiv \sum_{\nu = 1}^{\infty} \left( E_{\nu} - E_{0} \right)^{n} \left| \big< \Psi_{\nu} \big| \mathcal{O} \big| \Psi_{0} \big> \right|^{2}, 
\label{Eq-A6}
\end{equation} 
where $\big| \Psi_{\nu} \big>$ indicates the $\nu$-th energy-eigenstate with the energy $E_{\nu}$ in the complete set, 
\begin{equation}
\sum_{\nu = 0}^{\infty} \big| \Psi_{\nu} \big> \big< \Psi_{\nu} \big| 
= 1. 
\label{Eq-A6-2}
\end{equation} 
As a result, we obtain the frequency $\Omega$ of the collective oscillation as 
\begin{equation}
\Omega 
= \sqrt{\frac{M_{3}}{M_{1}}} 
\label{Eq-A7}
\end{equation}
according to eqs.~(\ref{Eq-A4})-(\ref{Eq-A5-2}). This result is equivalent to that in the sum-rule method. 
\section{The Thomas-Fermi approximation \label{App-B}}
In the single-particle picture, the Wigner function in eq.~(\ref{Eq-44}) can be described as 
\begin{equation}
f(\vec{r}, \vec{p}) 
= \int d{\varepsilon}~ F(\varepsilon) D(\varepsilon) \mathfrak{f}(\vec{r}, \vec{p}; \varepsilon) 
\label{Eq-B1}
\end{equation}
with the Fermi distribution function 
\begin{equation}
F(\varepsilon) 
\equiv \frac{1}{e^{( \varepsilon - \mu ) / T} + 1} 
\label{Eq-B1-2}
\end{equation}
and density of states 
\begin{equation}
D(\varepsilon) 
\equiv \sum_{n} \delta{\left( \varepsilon - \varepsilon_{n} \right)}, 
\label{Eq-B3}
\end{equation}
where we introduce the single-particle Wigner function 
\begin{equation}
\mathfrak{f}(\vec{r}, \vec{p}; \varepsilon) 
\equiv \sum_{n} \delta_{\varepsilon, \varepsilon_{n}} \mathfrak{f}_{n}(\vec{r}, \vec{p}) 
\label{Eq-B2}
\end{equation}
and 
\begin{equation}
\mathfrak{f}_{n}(\vec{r}, \vec{p}) 
\equiv \int d{\vec{s}}~ e^{- i \vec{p} \cdot \vec{s}} \phi_{n}^{*}{\left( \vec{r} - \frac{\vec{s}}{2} \right)} \phi_{n}{\left( \vec{r} + \frac{\vec{s}}{2} \right)} 
\label{Eq-B2-2}
\end{equation}
with the single-particle wave-functions $\phi_{n}(\vec{r})$ and energy $\varepsilon_{n}$. 

In TFA, the Wigner function $f(\vec{r}, \vec{p})$ in eq.~(\ref{Eq-B1}) is semi-classically evaluated as 
\begin{equation}
D(\varepsilon) \mathfrak{f}(\vec{r}, \vec{p}; \varepsilon) 
\approx \delta{\left( \varepsilon - e_{\text{cl}}(\vec{r}, \vec{p}) \right)} 
\label{Eq-B4}
\end{equation}
and 
\begin{equation}
f(\vec{r}, \vec{p}) 
\approx F{\left( e_{\text{cl}}(\vec{r}, \vec{p}) \right)} 
\label{Eq-B5}
\end{equation}
with the classical energy $e_{\text{cl}}(\vec{r}, \vec{p})$. Here we keep the quantum statistics introduced in the Fermi distribution function $F(\varepsilon)$ in eq.~(\ref{Eq-B1-2}) and replace the quantum-mechanical density of states in eq.~(\ref{Eq-B3}) with that in classical mechanics in eq.~(\ref{Eq-B4}). The above result can also be obtained from the semi-classical $\hbar$ expansion~\cite{Parr}. 

By integrating the delta function in eq.~(\ref{Eq-B4}) in the whole phase-space, we can obtain the classical energy-level density owing to the normalization of the single-particle Wigner function $\mathfrak{f}_{n}(\vec{r}, \vec{p})$ in eq.~(\ref{Eq-B2-2}). 

In the $d$-dimensional system, it becomes 
\begin{equation}
D^{(d)}(\varepsilon) 
= \iint \frac{d{\vec{r}} d{\vec{p}}}{(2 \pi)^{d}}~ \delta{\left( \varepsilon - e_{\text{cl}}(\vec{r}, \vec{p}) \right)}, 
\label{Eq-B7-0}
\end{equation}
where $\vec{r} = \big( r_{1}, r_{2}, \dots, r_{d} \big)$ and $\vec{p} = \big( p_{1}, p_{2}, \dots, p_{d} \big)$. In particular, in the $d$-dimensional harmonic oscillator systems, 
\begin{equation}
e_{\text{cl}}(\vec{r}, \vec{p}) 
= \frac{1}{2} \sum_{j = 1}^{d} \left( p_{j}^{2} + \omega_{j}^{2} r_{j}^{2} \right), 
\label{Eq-B6}
\end{equation}
we obtain 
\begin{equation}
D^{(d)}(\varepsilon) 
= \frac{\varepsilon^{d - 1}}{(d - 1)!} \left( \prod_{j = 1}^{d} \omega_{j}^{-1} \right) \Theta{\left( \varepsilon \right)} 
\label{Eq-B7}
\end{equation}
with the Heaviside step function $\Theta{\left( x \right)}$. 

\begin{thebibliography}{10}
%
%
%
\bibitem{2007-01-TMTN} 
T. Maruyama and T. Nishimura: 
Phys. Rev. A {\bf 75} (2007) 033611. 
%
\bibitem{BECrv} 
For reviews, see 
A. S. Parkins and H. D. F. Walls: 
Phys. Rep. {\bf 303} (1998) 1; 
F. Dalfovo, S. Giorgini, L. P. Pitaevskii, and S. Stringari: 
Rev. Mod. Phys. {\bf 71} (1999) 463; 
W. Ketterle, D. S. Durfee, and D. M. Stamper-Kum: 
in {\it Bose-Einstein Condensation in Atomic Gases}, Proceedings of International School of Physics ``Enrico Fermi'', edited by M. Ingusico, S. Stringari and C. Wieman (IOS Press, Amsterdam, 1999); 
C. J. Pethik and H. Smith: 
{\it Bose-Einstein Condensation in Dilute Gases} 
(Cambridge University Press, Cambridge, 2002). 
%
\bibitem{nobel} 
E. A. Cornell and C. E. Wieman: 
Rev. Mod. Phys. {\bf 74} (2002) 875; 
W. Ketterle: 
Rev. Mod. Phys. {\bf 74} (2002) 1131. 
%
\bibitem{bec} 
M. H. Anderson, J. R. Ensher, M. R. Matthews, C. E. Wieman, and E. A. Cornell: 
Science {\bf 269} (1995) 198; 
K. B. Davis, M. O. Mewes, M. R. Andrews, N. J. van Druten, D. S. Durfee, D. M. Kurn, and W. Ketterle: 
Phys. Rev. Lett. {\bf 75} (1995) 3969. 
%
\bibitem{ferG-1} 
B. DeMarco and D. S. Jin: 
Science {\bf 285} (1999) 1703. 
%
\bibitem{ferG-2} 
S. R. Granade, M. E. Gehm, K. M. O'Hara, and J. E. Thomas: 
Phys. Rev. Lett. {\bf 88} (2002) 120405. 
%
\bibitem{BferM-1}
A. G. Tuscott, K. E. Strecker, W. I. McAlexander, G. B. Parridge, and R. G. Hullet: 
Science {\bf 291} (2001) 2570. 
%
\bibitem{BferM-2}
F. Schreck, L. Khaykovich, K. L. Corwin, G. Ferrari, T. Bourdel, J. Cubizolles, and C. Salomon: 
Phys. Rev. Lett. {\bf 87} (2001) 080403. 
%
\bibitem{BferM-3}
Z. Hadzibabic, C. A. Stan, K. Dieckmann, S. Gupta, M. W. Zwierlein, A. Gorlitz, and W. Ketterle: 
Phys. Rev. Lett. {\bf 88} (2002) 160401. 
%
\bibitem{BferM-4}
Z. Hadzibabic, S. Gupta, C. A. Stan, C. H. Schunck, M. W. Zwierlein, K. Dieckmann, and W. Ketterle: 
Phys. Rev. Lett. {\bf 91} (2003) 160401. 
%
\bibitem{BCSBECe} 
C. A. Regal, M. Greiner, and D. S. Jin: 
Phys. Rev. Lett. {\bf 92} (2004) 040403; 
M. Bartenstein, A. Altmeyer, S. Riedl, et al.: 
Phys. Rev. Lett. {\bf 92} (2004) 120401; 
M. N. Zwierlein, C. A. Stan, C. H. Schunck, et al.: 
Phys. Rev. Lett. {\bf 92} (2004) 120403. 
%
\bibitem{BCSBECt} 
For reviews, see 
S. Giorgini, L. P. Pitaevskii, and S. Stringari: 
Rev. Mod. Phys. 80 (2008) 1215. 
%
\bibitem{SY} 
T. Sogo and H. Yabu: 
Phys. Rev. A {\bf 66} (2002) 043611. 
%
\bibitem{lowdime-1} 
A. G\"{o}rlitz, J. M. Vogels, A. E. Leanhardt et al.: 
Phys. Rev. Lett. {\bf 87} (2001) 130402. 
%
\bibitem{lowdime-2} 
S. Dettmer, D. Hellweg, P. Ryytty et al.: 
Phys. Rev. Lett. {\bf 87} (2001) 160406. 
%
\bibitem{lowdime-3} 
F. Schreck, L. Khaykovich, K. L. Corwin, G. Ferrari, T. Bourdel, J. Cubizolles, and C. Salomon: 
Phys. Rev. Lett. {\bf 87} (2001) 080403. 
%
\bibitem{lowdime-4} 
T. St\"{o}ferle, H. Moritz, C. Schori, M. Kohl, and T. Esslinger: 
Phys. Rev. Lett. {\bf 92} (2004) 130403. 
%
\bibitem{lowdime-a1} 
H. Ott, J. Fortagh, G. Schlotterbeck, A. Grossmann, and C. Zimmermann: 
Phys. Rev. Lett. {\bf 87} (2001) 230401. 
%
\bibitem{lowdime-a2} 
W. H\"{a}nsel, P. Hommelhoff, T. W. H\"{a}nsch, J. Reichel: 
Nature {\bf 413} (2001) 498. 
%
\bibitem{lowdime-a3} 
J. H. Thywissen, M. Olshanii, G. Zabow et al.: 
Eur. Phys. J. D {\bf 7} (1999) 361. 
%
\bibitem{lowdime-a4} 
S. Schneider, A. Kasper, Ch. vom Hagen et al.: 
Phys. Rev. A {\bf 67} (2003) 023612. 
%
\bibitem{lowdimt} 
For review, see 
L. Pitaevskii and S. Stringari: 
{\it Bose-Einstein Condensation} 
(Clarendon Press, Oxford, 2003); 
D. S. Petrov, D. M. Gangardt, and G. V. Shlyapnikov: 
J. Phys. IV {\bf 116} (2006) 3. 
%
\bibitem{ClFer} 
G. M. Bruun: 
Phys. Rev. A {\bf 63} (2001) 043408. 
%
\bibitem{Kry} 
K. Goral, M. Brewczyk, and K. Rzazewski: 
Phys. Rev. A {\bf 67} (2003) 025601. 
%
\bibitem{ClSp} 
G. M. Bruun and B. R. Mottelson: 
Phys. Rev. Lett {\bf 87} (2001) 270403. 
%
\bibitem{sumF} 
L. Vichi and S. Stringari: 
Phys. Rev. A {\bf 60} (1999) 4734. 
%
\bibitem{scal} 
G. F. Bertsch: 
Nucl. Phys. A {\bf 249} (1975) 253; 
G. F. Bertsch and K. Stricker: 
Phys. Rev. C {\bf 13} (1976) 1312; 
D. M. Brink and R. Leobardi: 
Nucl. Phys. A {\bf 258} (1976) 285; 
T. Suzuki: 
Prog. Theor. Phys. {\bf 64} (1980) 1627. 
%
\bibitem{bohi} 
O. Bohigas, A. M. Lane, and J. Martorell: 
Phys. Rep. {\bf 51} (1979) 267. 
%
\bibitem{ToBe} 
T. Maruyama and G. F. Bertsch: 
Phys. Rev. A {\bf 73} (2006) 013610. 
%
\bibitem{FetterWalecka}
A. L. Fetter and J. D. Walecka: 
{\it Quantum theory of many-particle systems} 
(McGrawHill, NY, 1971). 
%
\bibitem{Nilsson} 
The precise structures of the single-particle energy-level densities are studied well in the Nilsson model for nuclear physics; See 
P. Ring and P. Schuck: 
{\it The Nuclear Many-Body Problem} 
(Springer, New York, 2000). 
%
\bibitem{footnote-1} 
In general, the QLD structures produce the multi-time-scale structures, where the coarse graining yields the QLD properties in the slowest time-scale with renormalization of the degrees of freedom associated with the faster time-scales. 
%
\bibitem{TM-01} 
T. Maruyama, H. Yabu, and T. Suzuki: 
Phys. Rev. A {\bf 72} (2005) 013609. 
%
\bibitem{TM-02} 
T. Maruyama and G. F. Bertsch: 
Phys. Rev. A {\bf 77} (2008) 063611. 
%
\bibitem{Parr} R. G. Parr and W. Yang Weitao: 
{\it Density-functional Theory of Atoms And Molecules} 
(Oxford University Press, NewYork, 1989). 
%






\end{thebibliography}
\end{document}